\UseRawInputEncoding
\documentclass[superscriptaddress, twocolumn, amsmath, amssymb, aps,pra, notitlepage,longbibliography,10pt]{revtex4-1} %preview twocolumn
\usepackage{graphicx,graphics,epsfig,subfigure,times,bm,bbm,amssymb,amsmath,amsfonts,amsthm,mathrsfs,MnSymbol,physics}
\usepackage[matrix,frame,arrow]{xypic}
\usepackage[normalem]{ulem}
\usepackage{slashed}
\usepackage{dcolumn}
\usepackage{tabularx}
\usepackage{array}
\usepackage{adjustbox}
\usepackage{amsopn}
\usepackage{color}
\usepackage[usenames,dvipsnames,svgnames,table]{xcolor}
\usepackage[english]{babel}
\usepackage{verbatim}
\definecolor{darkblue}{rgb}{0.0,0.0,0.3}
\usepackage[colorlinks=true,
            linkcolor=red,
            urlcolor= darkblue,
            citecolor=blue]{hyperref}

\usepackage{cleveref}

\usepackage{titlesec}
\titleclass{\subsubsubsection}{straight}[\subsection]
\newcounter{subsubsubsection}[subsubsection]
\renewcommand\thesubsubsubsection{\thesubsubsection.\arabic{subsubsubsection}}
\titleformat{\subsubsubsection}
  {\centering\normalfont\normalsize\itshape} 
  {\thesubsubsubsection}             
  {1em}                              
  {}                                 
\titlespacing{\subsubsubsection}
  {0pt}
  {3.25ex plus 1ex minus .2ex}
  {1.5ex plus .2ex}                
\newcommand{\bea}{\begin{eqnarray}}
\newcommand{\eea}{\end{eqnarray}}

\begin{document}

%\title{Dynamical transitions in real-space $PT$- and $APT$-symmetric non-Hermitian SSH models under pure dephasing}% Force line breaks with 
\title{Initial-state-dependent dephasing effect in non-Hermitian Su-Schrieffer-Heeger models}

\author{Tao Min}
\affiliation{Department of Physics, Institute for Quantum Science and Technology, Shanghai Key Laboratory of High Temperature Superconductors, International Center of Quantum and Molecular Structures, Shanghai University, Shanghai, 200444, China}
\author{Junjie Liu}
\email{jj\_liu@shu.edu.cn}
\affiliation{Department of Physics, Institute for Quantum Science and Technology, Shanghai Key Laboratory of High Temperature Superconductors, International Center of Quantum and Molecular Structures, Shanghai University, Shanghai, 200444, China}

\begin{abstract}
Understanding the dynamical evolution of non-Hermitian systems under external dissipation is essential for devising realistic non-Hermitian quantum devices that surpass the capabilities of their Hermitian counterparts. Dephasing, a major realistic dissipation, has an impact on non-Hermitian systems that remains largely unexplored. Here, we focus on finite-sized non-Hermitian Su-Schrieffer-Heeger (SSH) lattice models with alternating gain and loss in real space and examine the dynamical evolution of the trace distance under pure dephasing. By tuning system parameters, this model supports phases with either parity-time or anti-parity-time symmetries, enabling us to explore the dephasing effect across different non-Hermitian symmetry regimes within a single platform. While the trace distance exhibits distinct dynamical behaviors across the different phases in the absence of dephasing, its response to dephasing is largely symmetry-independent but instead initial-state dependent. By varying the initial states, we find that dephasing affects the trace distance in two qualitatively different ways. For initial states that differ only in their coherences, dephasing accelerates the decay of the trace distance, destroying the phase information in much the same way as in Hermitian systems. In contrast, for initial states that differ only in their populations, strong dephasing can partially or completely stabilize the trace distance over a wide transient regime before its eventual decay. We attribute these metastable stabilization phenomena to the alternating gain and loss in the system and confirm their absence in Hermitian counterparts. Our findings reveal the intricate dynamical response of non-Hermitian SSH models to dephasing, thereby facilitating their potential applications.
\end{abstract}

\maketitle
\date{\today}

\section{Introduction}

Non-Hermitian quantum mechanics has emerged as a transformative framework, extending the boundaries of traditional quantum theory beyond the constraints of Hermiticity~\cite{bender.2007.rpp,Ganainy.18.NP,Ashida.20.AP,Bergholtz.21.RMP,Okuma.23.ARCMP,Bender.24.RMP}. Unlike their Hermitian counterparts, non-Hermitian systems exhibit unique spectral features, most notably the appearance of exceptional points (EPs) where eigenvalues and eigenvectors coalesce~\cite{Heiss.04.JPA,mandal.2021.prl,miri2019exceptional,Ozdemir.19.NM,Minganti.19.PRA,Ding.22.NRP,Xue.26.PRL}. By marking the points of spectral reconfiguration, EPs usually define the transition points between symmetry-unbroken and symmetry-broken regimes. This spectral flexibility facilitates the emergence of unconventional symmetries such as parity-time ($\mathcal{PT}$) and anti-parity-time ($\mathcal{APT}$) symmetries~\cite{Bender.24.RMP,Bender.98.PRL,Bender.99.JMP,Makris.08.PRL,peng.2016.np,li2024experimental,konotop2018odd,choi.2018.nc,El-Ganainy.18.NP,Xiao.19.PRL,Ozdemir.19.NM,ke2019topological,li2019anti,Fang.21.CP,Almanakly.26.PRL}. A typical approach to constructing non-Hermitian systems with certain symmetries is the introduction of gain and loss into conventional Hermitian systems \cite{Longhi.15.SP,Mochizuki.16.PRA,Shi.16.NC,Auregan.17.PRL,Xiao.17.NP,Rivet.18.NP,Luo.19.PRL,Zhao.19.S,Song.19.PRL,Scheibner.20.PRL,LiuS.20.PRA,Jiang.24.PRB,Pu.26.PRL}. Serving as a quintessential example of this paradigm, the integration of non-Hermiticity into the Su-Schrieffer-Heeger (SSH) model~\cite{Su.80.PRB} has emerged as a focal point of contemporary research~\cite{Rudner.09.PRL,Schomerus.13.OL,Esaki.11.PRB,Yuce.15.PLA,Zhu.14.PRA,Zeuner.15.PRL,Poli.15.NC,Weimann.17.NM,Kunst.18.PRL,Yao.18.PRL,asboth2016short,meier2016observation,chen2019finite,obana2019dimensional,lieu2018topological,hu2023anti,wu2021topology,li2024experimental,xue2022non,Pyrialakos.22.PRL,Jangjan.24.PRB}. Particularly, recent studies have further extended non-Hermitian SSH model to incorporate long-range hopping~\cite{Yao.18.PRL}, higher periodicity~\cite{li2014topological,he2021non}, nonlinearity~\cite{Yuce.26.PRB,Li.26.PRB} and even many-body interactions \cite{di2016two,yahyavi2018variational}.

The exotic static properties of non-Hermitian systems suggest their potential for unique functionalities in quantum information-processing~\cite{Zhang.12.PRA,ren.2022.np,Abbasi.22.PRL,Arkhipov.24.PRL}. Fully harnessing this potential, however, necessitates a shift in focus toward their dynamical behaviors~\cite{LiuW.21.PRL,Liang.22.PRL,Xiao.19.PRL,Kawabata.17.PRL,Wen.20.NPJQI,Fang.21.CP,fang2022entanglement,DingYL.2022.pra,ma2025dynamical,akram.2023.sr}. Given that experimentally realized non-Hermitian systems are typically effective systems that remain open to surrounding environments~\cite{Ozdemir.19.NM,ren.2022.np,chenwj.2021.prl,chenwj.2022.prl,Almanakly.26.PRL}, the impact of environmental dissipation on their dynamical properties must be carefully characterized~\cite{ma2025dynamical,Longhi.24.LSA,sun.2023.pra,Wang.26.A}. In practical quantum information processing, dephasing is a ubiquitous and typically detrimental source of noise that leads to the loss of information. While dephasing usually accelerates the decay of quantum information in Hermitian systems, its role in non-Hermitian systems might be fundamentally different due to the nontrivial spectral structure. Recent studies have hinted that the combination of non-Hermiticity and dissipation can lead to counterintuitive effects, such as decoherence-induced EPs~\cite{chenwj.2022.prl} and dephasing-induced protection~\cite{ma2025dynamical}. Understanding how the dynamical signatures of $\mathcal{PT}$ and 
$\mathcal{APT}$ symmetries persist or evolve under dephasing is therefore crucial for both fundamental physics and the design of robust non-Hermitian quantum devices.

In this work, we investigate non-Hermitian SSH lattice models with alternating gain and loss in real space, subject to local pure dephasing. Crucially, by simply tuning system parameters, this non-Hermitian lattice model hosts three distinct phases: the $\mathcal{PT}$-unbroken, $\mathcal{APT}$-unbroken and symmetry broken phases. This versatility allows us to explore the interplay between various non-Hermitian symmetries and dephasing within a unified platform. In contrast to previous studies on non-Hermitian SSH models, which typically analyze the corresponding Bloch Hamiltonian in momentum space under the assumption of a perfect periodicity in real space--a condition often challenging to realize in real-space experiments, we work directly with finite-sized lattice systems in real space. Consequently, our results correspond directly to the system under consideration, thereby facilitating direct experimental verification.

Serving as benchmarks for isolating the dynamical responses of non-Hermitian SSH models to dephasing, we first characterize the dynamical signatures of three phases as imprinted on the evolution of the trace distance~\cite{nielsen2001quantum}, which is an information-theoretic quantity that quantifies the information flow~\cite{Kawabata.17.PRL,Xiao.19.PRL,ma2025dynamical}. We show that the dynamics of the trace distance exhibits persistent oscillation, damped oscillation and superpositions of exponential decays in the $\mathcal{PT}$-unbroken, symmetry broken and $\mathcal{APT}$-unbroken phases, respectively. These distinct behaviors originate from the differences in the Hamiltonian spectrum across the different phases.

Turning to open non-Hermitian SSH models subject to dephasing, we adopt a generalized quantum Lindblad master equation~\cite{ma2025dynamical,yuan2020steady,PhysRevLett.132.110402,sun.2023.pra} to describe the dynamics of the normalized system density matrix. Interestingly, we find that the basic trends of the trace distance evolution under varying dephasing strengths remain almost the same across different phases and are sensitive only to the initial state. By varying the initial states within the single-excitation subspace, we identify two basic effects of dephasing: one in which dephasing merely accelerates the decay of the trace distance, playing a detrimental role similar to its impact in Hermitian systems, and another in which dephasing transiently stabilizes the trace distance in the strong-dephasing limit over a wide time window. The latter occurs only for incoherent initial states that differ solely in their populations.

Notably, this dephasing-induced transient stabilization takes two forms depending on the initial state: partial stabilization, where the trace distance approaches a finite value smaller than its initial value under strong dephasing, and complete stabilization, where the dynamical evolution of the trace distance is completely frozen in the strong-dephasing limit, so that it remains at its initial value. In the long-time limit, however, both types of stabilization vanish and the trace distance ultimately decays to zero. By analyzing the equations of motion for the density matrix elements in the strong-dephasing limit, we attribute the initial-state dependence of the dephasing response to the alternating gain and loss in real space, since initial excitations starting from gain versus loss sites lead to distinct dynamics.
%Moreover, by examining the trend toward complete stabilization upon increasing the dephasing strength, we identify an interesting type of dynamical response of the non-Hermitian SSH model in the $\mathcal{APT}$-symmetry-unbroken phase: a continuous suppression {\color{red}of the intrinsic decay of the trace distance} upon increasing the dephasing strength is observed, which is absent in other phases and in its Hermitian counterpart. We elucidate this continuous suppression via the Liouvillian gap, thereby marking a beneficial role of dephasing in protecting {\color{red}state distinguishability}. 

We note that Ref.~\cite{ma2025dynamical} also reported a dephasing-induced stabilization effect in non-Hermitian systems. However, the present study differs from and significantly extends that work in several key aspects. First, Ref.~\cite{ma2025dynamical} demonstrated this effect only in the $\mathcal{APT}$-symmetry-unbroken phase of non-interacting non-Hermitian qubit arrays. In stark contrast, we find that dephasing-induced transient stabilization occurs in all three symmetry regimes of the interacting non-Hermitian SSH lattice: the $\mathcal{PT}$-symmetry-unbroken, symmetry-broken, and $\mathcal{APT}$-symmetry-unbroken phases. Second, we uncover a pronounced initial-state dependence of the dephasing effect. Third, our system exhibits a distinct transient partial stabilization effect. We emphasize that these features, which are entirely absent in Ref.~\cite{ma2025dynamical}, originate directly from the alternating gain and loss structure inherent in the non-Hermitian SSH lattice. Collectively, these differences establish our work as an independent and conceptually distinct contribution to the understanding of dephasing effects in non-Hermitian systems, offering insights that go well beyond those reported in Ref.~\cite{ma2025dynamical}.

The structure of the paper is organized as follows. In Sec.~\ref{TWO}, we introduce a minimal four-site non-Hermitian SSH lattice model with alternating gain and loss in real space and determine its phase diagram through a comprehensive symmetry analysis. We also present the evolution dynamics of the trace distance in the absence of dephasing for later comparison. In Sec.~\ref{THREE}, we explore the dynamical behaviors of the trace distance under pure dephasing. By combining analytical equation-of-motion analysis with extensive numerical simulations, we provide an in-depth analysis of the initial-state-dependent dynamical response to dephasing. In Sec.~\ref{sec:multi_site}, we further demonstrate the robustness of our findings in a multi-site non-Hermitian SSH lattice. Finally, we summarize the study in Sec.~\ref{sec:conclusion}.

\section{Non-Hermitian SSH model: Phase diagram and closed-system dynamics}\label{TWO}

\begin{figure}[bth!]
 \centering
 \includegraphics[width=1\columnwidth]{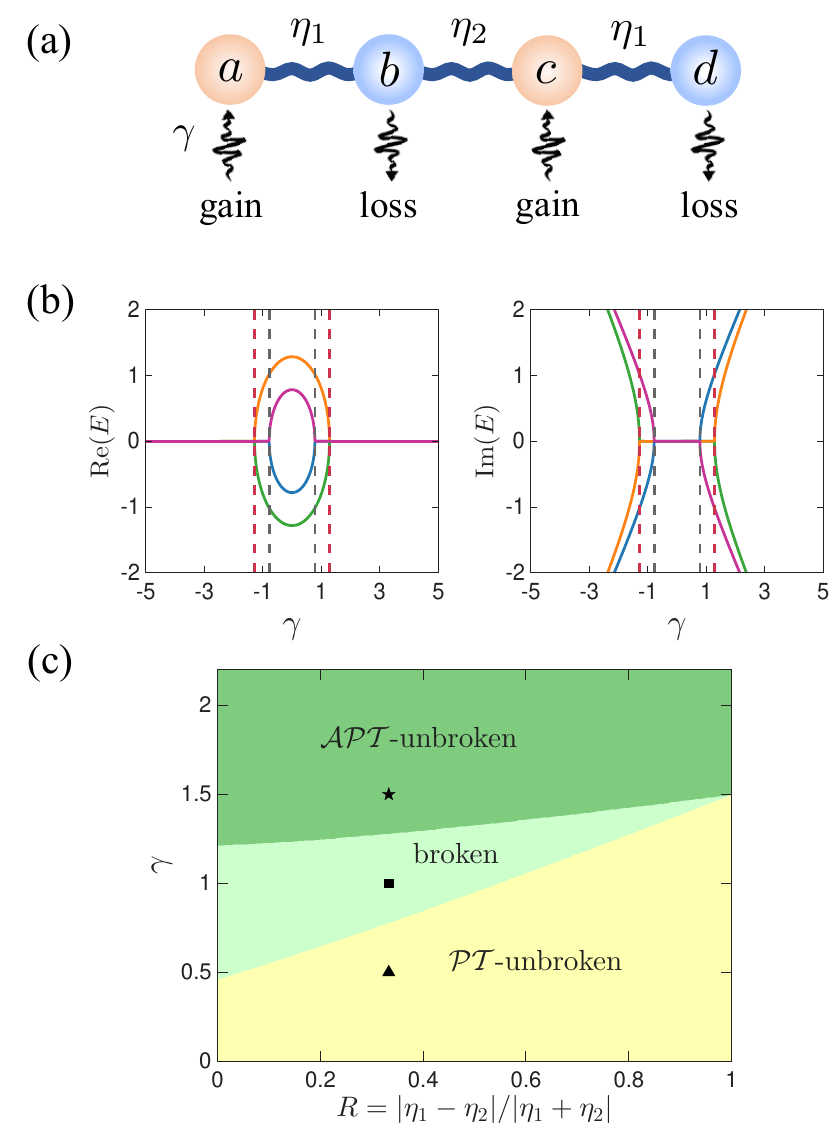}
\caption{(a) Sketch of the minimum four-site non-Hermitian SSH model with alternating hopping strengths $\eta_1$ and $\eta_2$, and alternating gain and loss processes measured by strength $\gamma$. (b) Real (left panel) and imaginary (right panel) parts of the energy eigenvalues as a function of $\gamma$ with $\eta_1=1$ and $\eta_2=0.5$. The inner grey and outer red vertical dashed lines indicate the locations of the exceptional points at $\gamma = \pm \sqrt{u_-}$ and $\gamma =\pm \sqrt{u_+}$, respectively, where $u_{\pm} = (2\eta_1^2 + \eta_2^2 \pm \eta_2\sqrt{4\eta_1^2 + \eta_2^2})/2$. 
(c) Phase diagram in the $(\gamma, R)$ parameter space for $\eta_1 > \eta_2$, where $R = |\eta_1 - \eta_2| / (\eta_1 + \eta_2)$. The yellow, light green, and dark green shaded regions correspond to the $\mathcal{PT}$-unbroken, symmetry broken, and $\mathcal{APT}$-unbroken phases, respectively. The triangle, square, and star markers at $R = 1/3$ and $\gamma \in \{0.5, 1.0, 1.5\}$, respectively, designate the specific parameter sets selected for later simulations in the region with $\eta_1 > \eta_2$.}\label{1}
\end{figure}

\subsection{Model setup}
We consider a minimal one-dimensional non-Hermitian SSH lattice model that is not only simple to analyze and amenable to analytical treatment but also exhibits rich dynamical behaviors, allowing us to demonstrate the essential physics underlying the interplay between non-Hermiticity and dephasing. This minimal model, schematically illustrated in Fig.~\ref{1}~(a), consists of four sites and is described by the Hamiltonian
\begin{align}
     \hat{H} =& \eta_1(\hat{a}^\dagger \hat{b} + \hat{c}^\dagger \hat{d} + \rm{H.c.}) + \eta_2(\hat{b}^\dagger \hat{c} + \rm{H.c.}) \nonumber\\
     &+ i\gamma(\hat{a}^\dagger \hat{a} - \hat{b}^\dagger \hat{b} + \hat{c}^\dagger \hat{c} - \hat{d}^\dagger \hat{d}),
     \label{erci}
\end{align}
where $\hat{a}, \hat{b}, \hat{c}$ and $\hat{d}$ denote the annihilation operators (with the corresponding creation operators implied) of the first, second, third and fourth sites (see Fig.~\ref{1}~(a)), respectively. `H.c.' is short for Hermitian conjugate, and $\eta_{1,2}$ represent the alternating inter-site coupling strengths. Unless otherwise stated, we set $\eta_1\neq \eta_2$. We consider balanced gain and loss, with a single parameter $\gamma$ controlling the strength of the alternating gain and loss processes applied to the sites and thus quantifying the degree of non-Hermiticity. Throughout this study, we consider open boundary conditions.

Notably, the minimal model we consider functions as a fundamental unit cell for constructing extended multi-site non-Hermitian SSH lattice models~\cite{lieu2018topological,Weimann.17.NM,hu2023anti,chen2019finite}. In Sec.~\ref{sec:multi_site}, we will use this extension strategy to study a multi-site system, but prior to that section we focus on this minimal system. In contrast to existing studies that typically analyze the Bloch Hamiltonian in momentum space that inherently requires an infinite number of unit cells in real space, we work directly with finite-sized non-Hermitian SSH lattice models in real space. This choice offers two benefits. First, the system is amenable to implementation in current finite-sized experimental platforms. Second, the obtained results accurately reflect the properties of the specific real-space system, which is also important for experimental verification.

We restrict our analysis to the single-excitation subspace spanned by the Fock states $\{|S_1\rangle, |S_2\rangle, |S_3\rangle, |S_4\rangle\}$, where $|S_1\rangle = |1000\rangle$, $|S_2\rangle = |0100\rangle$, $|S_3\rangle = |0010\rangle$ and $|S_4\rangle = |0001\rangle$, with $|S_n\rangle$ representing an excitation on the $n$th site and all other sites empty. In this basis, the Hamiltonian in Eq. (\ref{erci}) is explicitly given by the following $4\times 4$ matrix
\begin{align}
    \hat{H} = \begin{pmatrix}
    i\gamma & \eta_1 & 0 & 0 \\
    \eta_1 & -i\gamma & \eta_2 & 0 \\
    0 & \eta_2 & i\gamma & \eta_1 \\
    0 & 0 & \eta_1 & -i\gamma
    \end{pmatrix}.
    \label{eq1}
\end{align}

\subsection{Spectrum and phase diagram}
Before proceeding to the dynamics, we first analyze the spectrum and symmetry properties of the model. Using the matrix form in Eq.~\eqref{eq1}, the energy eigenvalues $E$ satisfy the characteristic equation $\det(\hat{H} - E \hat{I}) = 0$, which reduces to the following relation
\begin{equation}\label{eq:3}
\begin{split}
    E^4 &+ (2\gamma^2 - 2\eta_1^2 - \eta_2^2 )E^2 \\
    &+ (\gamma^4 - 2\gamma^2\eta_1^2 - \gamma^2\eta_2^2 + \eta_1^4 ) = 0.
\end{split}
\end{equation}
Introducing auxiliary variables $u_{\pm} = (2\eta_1^2 + \eta_2^2 \pm \eta_2\sqrt{4\eta_1^2 + \eta_2^2})/2$, the energy eigenvalues (spectrum) are given by ($m=1,2,3,4$)
\begin{equation}\label{eq:Epm}
    \{E_{m}\}=\{\pm\sqrt{u_{\pm} - \gamma^2}\}.
\end{equation}
It is evident that the system possesses Hamiltonian exceptional points (EPs) at $|\gamma| = \sqrt{u_-}$ and $|\gamma| = \sqrt{u_+}$. Since $u_-<u_+$, we infer from Eq.~(\ref{eq:Epm}) that the spectrum is purely real when $\gamma^2 \leq u_-$. Increasing $\gamma^2$ such that $u_- < \gamma^2 < u_+$ yields a complex spectrum where one pair of eigenvalues is real and the other is imaginary. Increasing $\gamma^2$ further such that $\gamma^2 \geq u_+$ results in a purely imaginary spectrum. A set of energy eigenvalues is presented in Fig.~\ref{1}~(b), which clearly illustrates the spectral transitions upon increasing $\gamma^2$.

Whether the spectrum is complex, purely real, or purely imaginary depends on the underlying symmetry of the non-Hermitian system. Interestingly, the non-Hermitian SSH model we consider respects both $\mathcal{PT}$ and $\mathcal{APT}$ symmetries. Denoting the $\mathcal{PT}$ symmetry operator as the combined operation of spatial inversion $\hat{\sigma}_x \otimes \hat{\sigma}_x$ and complex conjugation $\mathcal{K}$ (i.e., $i \to -i$)~\cite{hu2023anti}, we can show that $[\hat{H}, \mathcal{PT}] = 0$ (see details in Appendix~\ref{a:1}). When the $\mathcal{PT}$ symmetry is unbroken, the spectrum is purely real~\cite{Bender.98.PRL}. Analogously, we introduce the $\mathcal{APT}$ operator as $\mathcal{APT} = \operatorname{diag}(1, -1, 1, -1)\mathcal{K}$~\cite{PhysRevLett.126.215302,hu2023anti}, with which we can show the anti-commutation relation $\{\hat{H}, \mathcal{APT}\} = 0$ holds (see  Appendix~\ref{a:1}). When the $\mathcal{APT}$ symmetry is unbroken, we instead have a purely imaginary spectrum~\cite{fang2022entanglement}.

Based on the relation between the spectrum and non-Hermitian symmetry, we immediately know that the non-Hermitian SSH model supports three phases: (i) a $\mathcal{PT}$-unbroken phase with a purely real spectrum, (ii) a symmetry broken phase with a complex spectrum, and (iii) a $\mathcal{APT}$-unbroken phase with a purely imaginary spectrum. By scanning the parameter space, we can obtain a phase diagram for the model. A representative phase diagram is shown in Fig.~\ref{1}~(c) for the case $\eta_1>\eta_2$. We have verified numerically that a similar phase diagram also exists for $\eta_1<\eta_2$.

In addition to the $\mathcal{PT}$- and $\mathcal{APT}$-symmetry transitions discussed above, the model also undergoes a topological transition at $\eta_1 = \eta_2$. Since both the hopping pattern $(\eta_1, \eta_2, \eta_1, \eta_2, \dots)$ and the gain-loss pattern $(+, -, +, -, \dots)$ have period two, the bulk is described by the two-site-unit-cell Bloch Hamiltonian $H(k) = i\gamma\sigma_z + h(k)\sigma_+ + h^*(k)\sigma_-$, with $h(k) = \eta_1 + \eta_2 e^{ik}$~\cite{Weimann.17.NM}. To characterize the topology, we adopt the winding number defined in Eq.~(1) of the Supplemental Material of Ref.~\cite{Weimann.17.NM}, which in our notation reads
\begin{equation}
w_h = \frac{1}{2\pi i} \int_0^{2\pi} q(k)^{-1} \frac{\partial}{\partial k} q(k) \, dk,
\end{equation}
where
\begin{equation}
q(k) = \frac{\eta_1 + \eta_2 e^{ik}}{\sqrt{|\eta_1 + \eta_2 e^{ik}|^2 - \gamma^2}}.
\end{equation}
In the $\mathcal{PT}$-unbroken phase, the square-root factor in $q(k)$ is real and positive and thus does not affect the phase of $q(k)$. The winding number therefore reduces to the $\gamma$-independent winding of $h(k)$, yielding $w_h = \Theta(\eta_2 - \eta_1)$, where $\Theta$ is the Heaviside step function: $w_h = 1$ for $\eta_1 < \eta_2$ and $w_h = 0$ for $\eta_1 > \eta_2$~\cite{Weimann.17.NM}. The topological transition thus occurs at $\eta_1 = \eta_2$. While the non-Hermitian term $\pm i\gamma$ modulates the eigenenergies, the localization properties and the existence of edge modes remain robust within the considered parameter range, provided that the $\mathcal{PT}$-unbroken phase is preserved~\cite{chen2019finite,Weimann.17.NM,lieu2018topological,PhysRevLett.126.215302}. Consequently, throughout this work we use $\eta_1 < \eta_2$ ($w_h = 1$) and $\eta_1 > \eta_2$ ($w_h = 0$) to denote the topologically nontrivial and trivial regimes, respectively. In Secs.~\ref{THREE} and \ref{sec:multi_site}, we will explicitly compare results in both regimes to explore whether topological properties affect the dephasing effects in finite-sized SSH lattices. Our results will indicate, however, that no clear topological effect emerges in finite-sized chains, as the trace distance in the strong-dephasing limit exhibits similar dynamical trends in both regimes. This is expected, since the winding number, defined via the Bloch Hamiltonian, strictly applies to an infinite chain.

\begin{figure}[t!]
    \centering
    \includegraphics[width=1\linewidth]{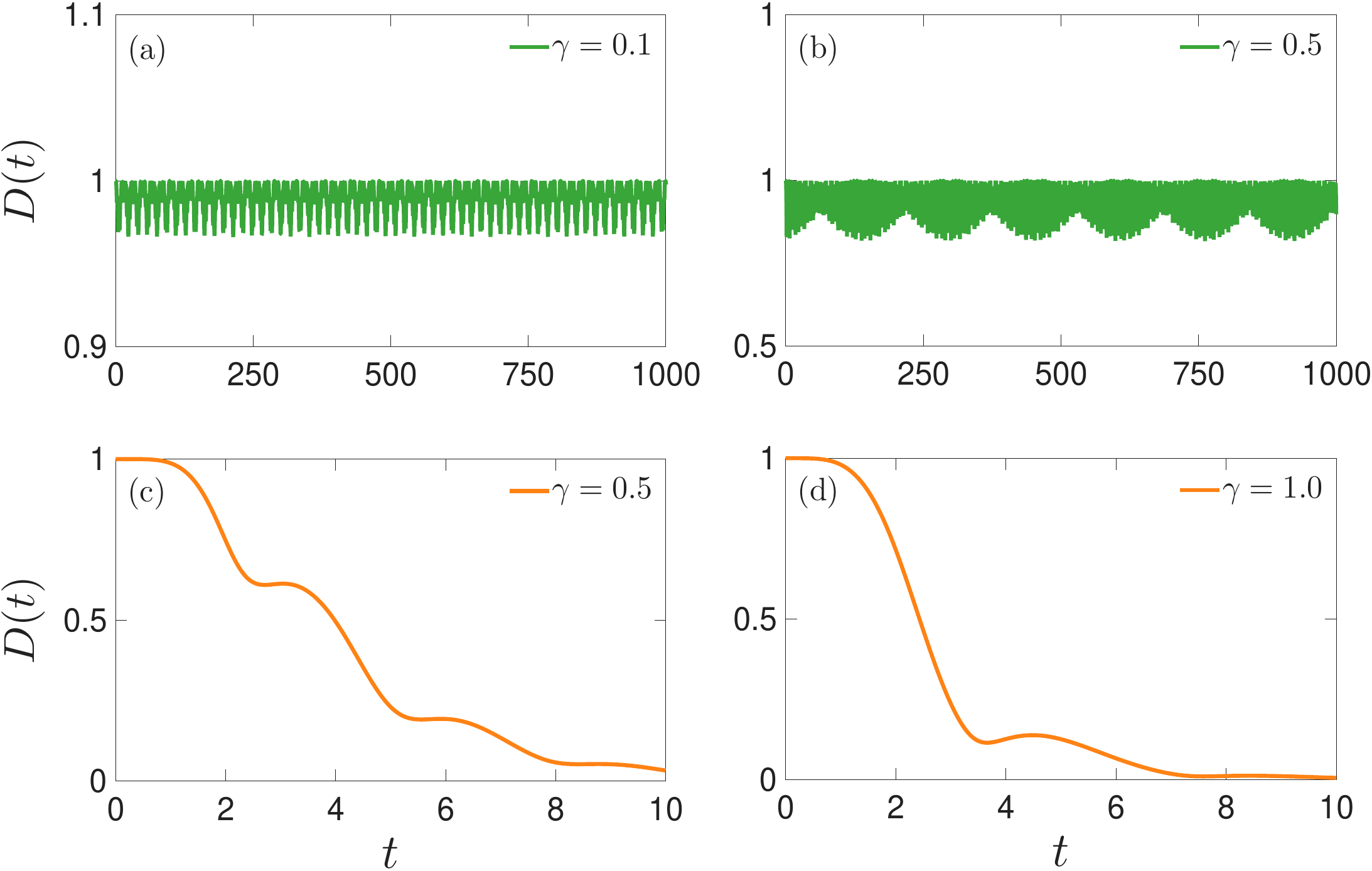}
    \caption{Dynamical evolution of the trace distance $D(t)$ in the closed non-Hermitian system. Upper panel: $\mathcal{PT}$-unbroken phase. Lower panel: symmetry broken phase. (a) and (c) correspond to $\eta_1 = 0.5$ and $\eta_2 = 1$. (b) and (d) correspond to $\eta_1 = 1$ and $\eta_2 = 0.5$. 
    The trace distance is computed for two initial conditions $\hat{\rho}_1(0)=|S_1\rangle\langle S_1|$ and $\hat{\rho}_2(0)=|S_3\rangle\langle S_3|$.}
    \label{B1}
\end{figure}

\subsection{Closed-system dynamics}
We now consider dynamics in the closed non-Hermitian SSH model which can provide benchmarks for isolating the dephasing effects. For closed non-Hermitian systems, the time evolution of the normalized density matrix $\hat{\rho}(t)$ is governed by the following equation \cite{ma2025dynamical,fang2022entanglement,Xiao.19.PRL,brody2012mixed,cao2023statistical,du2022physics,yuan2020steady,mao2024diagnosing}
\begin{equation}
    \hat{\rho}(t) = \frac{e^{-i\hat{H}t}\hat{\rho}(0)e^{i\hat{H}^{\dagger}t}}{\mathrm{Tr}\left[e^{-i\hat{H}t}\hat{\rho}(0)e^{i\hat{H}^{\dagger}t}\right]},
    \label{eq:closed_dynamics}
\end{equation}
where $\hat{H}$ is the non-Hermitian Hamiltonian, $\hat{\rho}(0)$ is the initial state. To reveal connections between Hamiltonian's spectrum and dynamical trends in closed systems, we consider the spectral decomposition of the density matrix in the biorthogonal eigenbasis of $\hat{H}$. By expanding the initial state in terms of the right eigenstates $\{|\phi_m\rangle\}$ and left eigenstates $\{\langle\chi_m|\}$ of $\hat{H}$, the density matrix at finite time $t$ can be exactly expressed as \cite{ma2025dynamical}
\begin{equation}
    \hat{\rho}(t) = \frac{\sum_{m,n} \rho_{mn} e^{-i(\lambda_m - \lambda_n)t} e^{(\Gamma_m + \Gamma_n)t} |\phi_m\rangle \langle \phi_n|}{\mathrm{Tr} \left[ \sum_{m,n} \rho_{mn} e^{-i(\lambda_m - \lambda_n)t} e^{(\Gamma_m + \Gamma_n)t} |\phi_m\rangle \langle \phi_n| \right]},
    \label{eq:rho_t_exact}
\end{equation}
where $\lambda_m$ and $\Gamma_m$ is the real and imaginary parts of energy eigenvalue $E_m$. The expansion coefficients $\rho_{mn} = \langle \chi_m | \hat{\rho}(0) | \chi_n \rangle / (\langle \chi_m | \phi_m \rangle \langle \phi_n | \chi_n \rangle)$. For convenience, we arrange the energy eigenvalues in descending order of their imaginary parts such that $\Gamma_1 \ge \Gamma_2 \ge \Gamma_3 \ge \Gamma_4$.  

Depending on the spectral properties in different phases, the density matrix generally exhibits three prototypical dynamical behaviors:

(i) In the $\mathcal{PT}$-unbroken phase, we have purely real spectrum such that $\{\Gamma_m = 0\}$ for all eigen modes. The dynamics are strictly dominated by phase accumulation terms $e^{-i(\lambda_m - \lambda_n)t}$, resulting in persistent, multi-frequency oscillations.

(ii) In the symmetry broken phase, we have a generally complex spectrum with nonvanishing real and imaginary parts of energy eigenvalues. This leads to damped oscillations, exhibiting a competition between persistent phase modulation and exponential relaxation.

(iii) In the $\mathcal{APT}$-unbroken phase, we instead have a purely imaginary spectrum with $\{\lambda_m = 0\}$.  The time evolution is entirely governed by the exponential terms $e^{(\Gamma_m + \Gamma_n)t}$. In the long-time limit, the decaying process is dominated by the leading-order term $e^{-\Delta \Gamma t}$ with the smallest decaying rate $\Delta\Gamma = \Gamma_1 - \Gamma_2$~\cite{ma2025dynamical}.

To illustrate the dynamical features in information-theoretic quantities, we specifically consider the evolution of the trace distance $D(t)$~\cite{nielsen2001quantum} throughout the study, 
\begin{equation}
D(t) \equiv \frac{1}{2} \mathrm{Tr} \left| \hat{\rho}_1(t)-\hat{\rho}_2(t) \right|,
\end{equation}
which quantifies state distinguishability and information flow~\cite{Kawabata.17.PRL,Xiao.19.PRL}. Here, $|\hat{A}|\equiv\sqrt{\hat{A}^{\dagger}\hat{A}}$. The trace distance vanishes, $D(t)=0$, when $\hat{\rho}_1(t)=\hat{\rho}_2(t)$, indicating that the two states are indistinguishable. Conversely, for orthogonal states, which are perfectly distinguishable, the trace distance attains its maximum value, $D(t)=1$.

Figure~\ref{B1} illustrate a set of dynamical results for the trace distance in both the $\mathcal{PT}$-unbroken phase (upper panel) and the symmetry broken phase (lower panel). As can be seen, the trace distance perfectly inherits the dynamical trends of the density matrix. In the $\mathcal{PT}$-unbroken phase [Figs.~\ref{B1}~(a) and (b)], $D(t)$ exhibits persistent oscillations. In the broken phase [Figs.~\ref{B1}~(c) and (d)], it shows damped oscillations. We have numerically checked the dynamical trends in these two phases are independent of the initial states.

\begin{figure}[b!]
    \centering
    \includegraphics[width=1\linewidth]{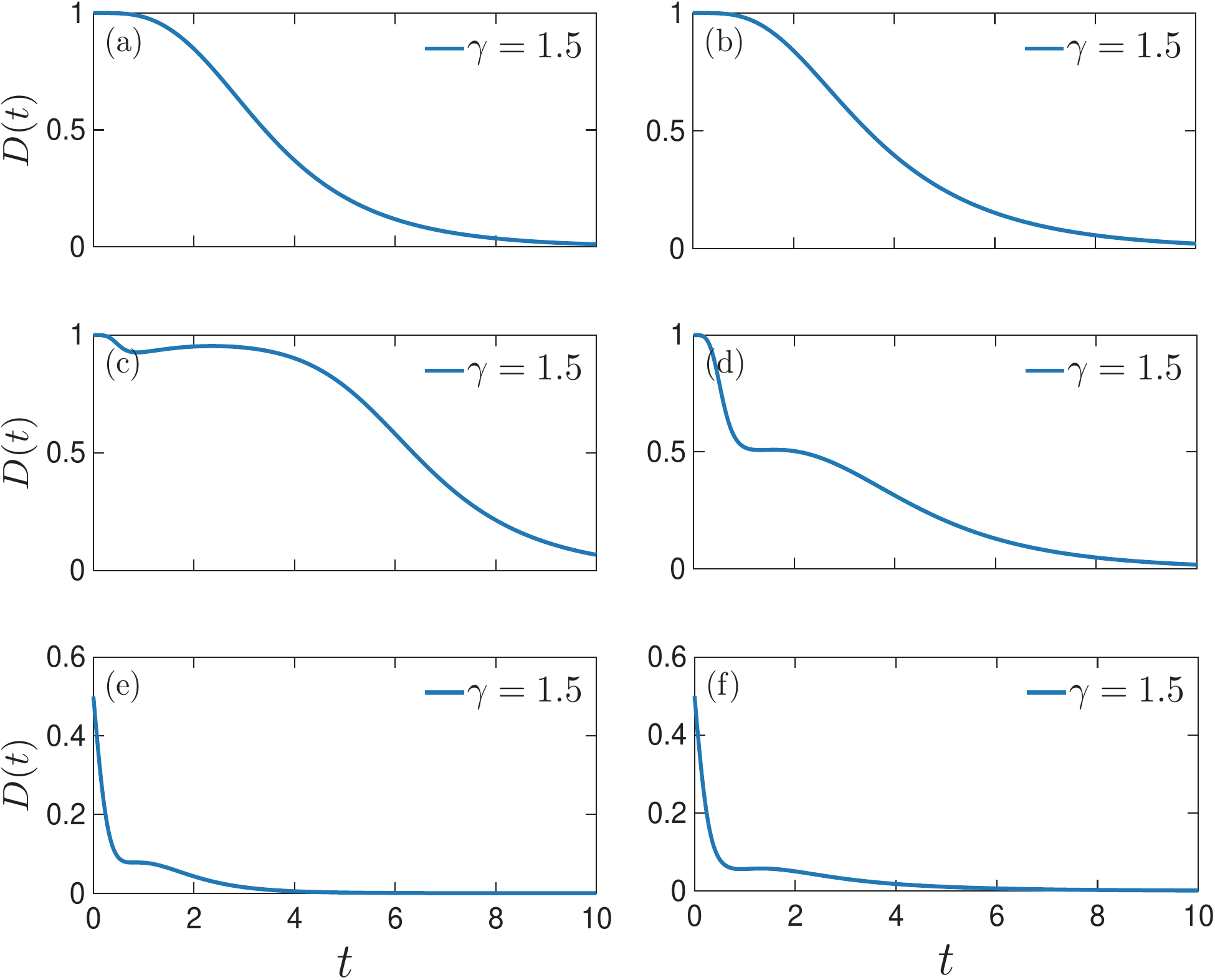}
    \caption{Time evolution of the trace distance $D(t)$ in the $\mathcal{APT}$-unbroken phase with $\gamma=1.5$ under different initial state pairs. Upper row [(a), (b)]: Initial states are $\hat{\rho}_1(0)=|S_1\rangle\langle S_1|$ and $\hat{\rho}_2(0)=|S_3\rangle\langle S_3|$. Middle row [(c), (d)]: Initial states are $\hat{\rho}_1(0)=|S_1\rangle\langle S_1|$ and $\hat{\rho}_2(0)=|S_2\rangle\langle S_2|$. Lower row [(e), (f)]: Initial states are $\hat{\rho}_1(0)=|S_1\rangle\langle S_1|$ and $\hat{\rho}_2(0)=0.5|S_1\rangle\langle S_1| + 0.5|S_2\rangle\langle S_2|$. Left panel: $\eta_1 = 0.5, \eta_2 = 1$. Right panel: $\eta_1 = 1, \eta_2 = 0.5$.}
    \label{A23}
\end{figure}

Interestingly, the dynamical trends of the trace distance in the $\mathcal{APT}$-unbroken phase exhibit rich decaying patterns that depend on the initial states. Representative results are shown in Figs.~\ref{A23} for three initial state pairs. From Figs.~\ref{A23} (a) and (b), we observe that under the initial pure state pair $\hat{\rho}_1(0)=|S_1\rangle\langle S_1|$ and $\hat{\rho}_2(0)=|S_3\rangle\langle S_3|$, the trace distance follows a monotonic exponential decay dominated by the leading-order term $e^{-\Delta \Gamma t}$. In contrast, for the different initial pure state pair $\hat{\rho}_1(0)=|S_1\rangle\langle S_1|$ and $\hat{\rho}_2(0)=|S_2\rangle\langle S_2|$, shown in Figs.~\ref{A23} (c) and (d), the trace distance exhibits a distinct non-monotonic evolution characterized by a prominent ``bump" or ``dip" before eventually decaying to zero. For comparison, we also consider an initial state pair that contains mixed initial state: $\hat{\rho}_1(0)=|S_1\rangle\langle S_1|$ and $\hat{\rho}_2(0)=0.5|S_1\rangle\langle S_1|+0.5|S_2\rangle\langle S_2|$, as shown in Figs.~\ref{A23} (e) and \ref{A23} (f). In this case, the trace distance drops rapidly at very short times, followed by a transient plateau before entering the asymptotic decaying regime. 

The rich decaying patterns arise from the structural inequivalence between sites in the non-Hermitian SSH lattice model due to alternating gain and loss, as well as from the initial-state-dependent superpositions of exponential decaying terms inherent in Eq.~(\ref{eq:rho_t_exact}), since different initial states yield different expansion coefficients. In the early stages of evolution, the rapidly decaying modes have not yet subsided. The superposition of these rapidly decaying modes, with their differing coefficients, can induce a transient overshoot or a plateau-like structure, depending on the specific combination of exponential terms $e^{(\Gamma_m+\Gamma_n)t}$. Only after these fast-decaying transient terms fully die out does the system enter the asymptotic regime, governed solely by the slowest decay rate, which restores the standard monotonic exponential decay. A detailed analysis is presented in Appendix~\ref{app:multi_exponential}.

These rich decaying patterns indicate that, while the long-time relaxation is universally dictated by the gap $\Delta \Gamma$, the transient short-time information dynamics in non-Hermitian SSH lattice models are highly sensitive to the spatial configuration of the initial states. These intrinsic closed-system features provide an essential baseline for our subsequent investigation of pure dephasing effects.

\section{Dynamical response to pure dephasing}
\label{THREE}
In this section, we turn to the effects of dephasing on the dynamical evolution of the trace distance in the model described by Eq.~(\ref{erci}). Conventionally, pure dephasing is known to be detrimental, as it inherently destroys quantum coherences. Indeed, we will first demonstrate that dephasing inevitably washes away any distinguishability encoded in the relative phases of the initial states. However, the situation becomes highly non-trivial when information is encoded in the spatial population distributions. Through both analytical analysis and extensive numerical simulations, we reveal an intriguing initial-state-dependent dephasing effect: depending on the initial population configuration, increasing dephasing can either accelerate the decay of the trace distance or paradoxically stabilize it at transient times.

\subsection{Quantum master equation}
To describe additional dephasing effects on non-Hermitian systems, we consider quantum master equation that describes the dynamical evolution of normalized density matrix $\hat{\rho}(t)$ \cite{ma2025dynamical,yuan2020steady,PhysRevLett.132.110402}
\begin{equation}
\begin{gathered}
    \frac{d}{dt}\hat{\rho}(t) = -i[\hat{H}\hat{\rho}(t) - \hat{\rho}(t)\hat{H}^{\dagger}] + \sum_k\mathcal{D}_k[\hat{\rho}(t)]\\
    - i\mathrm{Tr}[\hat{\rho}(t)(\hat{H}^{\dagger} - \hat{H})]\hat{\rho}(t),
    \label{eq:master}
\end{gathered}
\end{equation}
where $\hat{H}$ is the non-Hermitian Hamiltonian defined in Eq.~\eqref{eq1}. The term $\mathcal{D}_k[\hat{\rho}] = \gamma_k\left(\hat{L}_k\hat{\rho}\hat{L}_k^\dagger - \frac{1}{2}\{\hat{L}_k^\dagger\hat{L}_k,\hat{\rho}\}\right)$ represents the Lindblad dissipator capturing the external dissipation, where $\hat{L}_k$ and $\gamma_k$ denote the $k$-th jump operator and its corresponding decay rate, respectively. The last term on the right-hand-side of Eq. (\ref{eq:master}) denotes a correction term that ensures the conservation of probability~\cite{brody2012mixed}. In our model, we specifically focus on the effect of local pure dephasing, characterized by a uniform pure dephasing strength $\gamma_d$ across all sites (i.e., we set $\gamma_k = \gamma_d$ for all $k$) and local jump operators $\hat{L}_k = |{S_k}\rangle\langle {S_k}|$ ($k \in \{1,2,3,4\}$) in the single-excitation subspace.

Denote the Liouvillian superoperator $\mathcal{L}_0$ by $\mathcal{L}_0[\hat{\rho}] = -i[\hat{H}\hat{\rho} - \hat{\rho}\hat{H}^{\dagger}] + \sum_k\mathcal{D}_k[\hat{\rho}]$, with $e^{\mathcal{L}_0 t}$ its matrix exponential. Although $\mathcal{L}_0$ is linear, it is generally not trace preserving in non-Hermitian systems. Introducing the unnormalized operator $\tilde{\rho}(t) = e^{\mathcal{L}_0 t}\hat{\rho}(0)$, we note that the normalized density matrix is obtained as $\hat{\rho}(t) = \tilde{\rho}(t)/\operatorname{Tr}[\tilde{\rho}(t)]$. Indeed, this normalized state satisfies the nonlinear equation $\frac{d\hat{\rho}}{dt} = \mathcal{L}_0[\hat{\rho}] - \hat{\rho}\operatorname{Tr}(\mathcal{L}_0[\hat{\rho}])$, which is equivalent to Eq.~(\ref{eq:master}). Consequently, a stationary normalized state does not necessarily correspond to a zero eigenvalue of $\mathcal{L}_0$. Instead, it corresponds to a right eigenoperator $\hat{R}_1$ satisfying
\begin{equation}
    \mathcal{L}_0[\hat{R}_1]=\mu_1\hat{R}_1,
    \qquad
    \hat{\rho}_{\rm ss}
    =\frac{\hat{R}_1}{\operatorname{Tr}\hat{R}_1},
\end{equation}
provided that $\operatorname{Tr}\hat{R}_1\neq 0$.

To clarify the role of the Liouvillian spectrum of $\mathcal{L}_0$ in the relaxation dynamics of the system, let $\mathcal{L}_0[\hat{R}_j]=\mu_j\hat{R}_j$, with
$\mu_j=\mathcal{E}_j+i\alpha_j$. We sort these eigenvalues so that their real parts are in descending order, $\mathcal{E}_1 \geq \mathcal{E}_2 \geq \dots$. We have $\tilde{\rho}(t)=\sum_j c_j e^{\mu_j t}\hat{R}_j$, where the coefficients $c_j$ are the overlap between $\hat{\rho}(0)$ and the left eigenvectors of $\mathcal{L}_0$. Assuming that the dominant eigenvalue is unique and that $c_1\operatorname{Tr}\hat{R}_1\neq0$, one obtains at long times
\begin{equation}\label{xingshi}
    \hat{\rho}(t)
    =
    \frac{
    \hat{R}_1+
    \sum_{j>1}(c_j/c_1)e^{(\mu_j-\mu_1)t}\hat{R}_j
    }{
    \operatorname{Tr}\hat{R}_1+
    \sum_{j>1}(c_j/c_1)e^{(\mu_j-\mu_1)t}
    \operatorname{Tr}\hat{R}_j
    }.
\end{equation}
The common factor $e^{\mu_1t}$ cancels between the numerator and denominator. Consequently, the absolute value of $\mathcal{E}_1$ only describes the overall growth or decay of the unnormalized state, whereas the approach to the normalized steady state is controlled by the relative decay rate
\begin{equation}
    \Delta\mathcal{E}
    =\mathcal{E}_1-\max_{j\neq 1}\mathcal{E}_j.
\end{equation}
For the ordering
$\mathcal{E}_1\geq\mathcal{E}_2\geq\cdots$, this reduces to
$\Delta\mathcal{E}=\mathcal{E}_1-\mathcal{E}_2$, and the corresponding relaxation time is $\tau_o=\frac{1}{\Delta\mathcal{E}}$~\cite{ma2025dynamical}. When $\mathcal{L}_0$ defines a trace-preserving dynamics, we have
$\mu_1=0$ and this definition reduces to the conventional Liouvillian gap that is solely determined by $|\mathcal{E}_2|$. Given the above discussions, we refer to $\Delta\mathcal{E}$ as the Liouvillian gap for the non-Hermitian systems.

\begin{figure}[t!]
    \centering
    \includegraphics[width=1\linewidth]{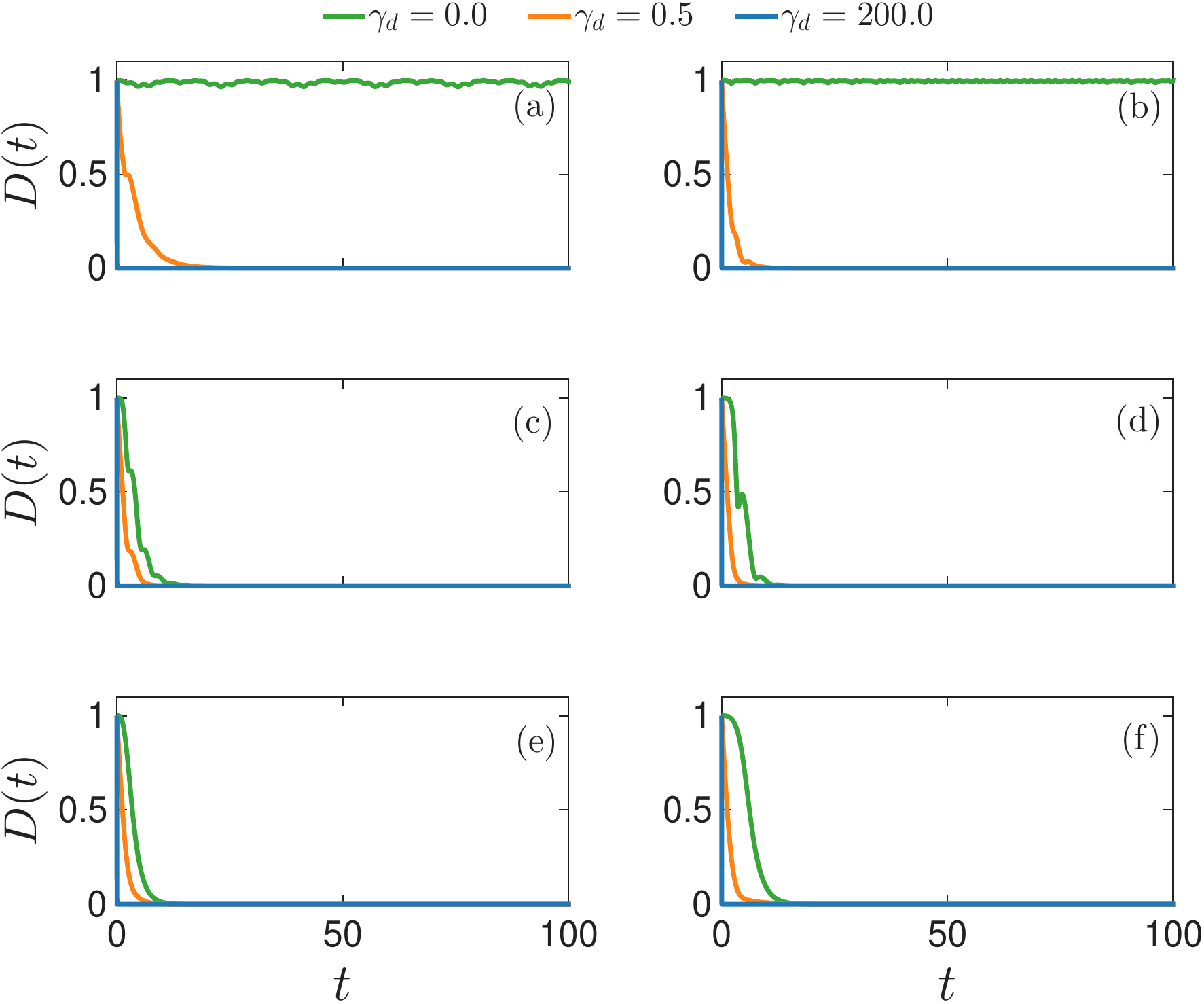}
    \caption{Dynamics of the trace distance $D(t)$ with varying dephasing strength $\gamma_d$ obtained by directly evolving Eq.~(\ref{eq:master}). (a) and (b) correspond to the $\mathcal{PT}$-unbroken phase for $\gamma = 0.1$ and $\gamma = 0.5$, respectively. (c) and (d) correspond to the symmetry-broken phase for $\gamma = 0.5$ and $\gamma = 1.0$, respectively. (e) and (f) correspond to the $\mathcal{APT}$-unbroken phase for $\gamma = 1.5$. Other parameters are $\eta_1 = 0.5$, $\eta_2 = 1$ for the left panel, and $\eta_1 = 1$, $\eta_2 = 0.5$ for the right panel. The initial states are $\hat{\rho}_1(0) = |\psi_1\rangle\langle \psi_1|$ and $\hat{\rho}_2(0)=|\psi_2\rangle\langle \psi_2|$ with $|\psi_1\rangle = (|S_1\rangle + |S_3\rangle)/\sqrt{2}$ and $|\psi_2\rangle = (|S_1\rangle - |S_3\rangle)/\sqrt{2}$.}
    \label{h1} 
\end{figure}

\subsection{Detrimental dephasing effect on coherence}
We first focus on the dephasing effect on relative-phase coherence. To this end, we consider a pair of initial states that differ only in their relative phase, $\hat{\rho}_1(0) = |\psi_1\rangle\langle \psi_1|$ and $\hat{\rho}_2(0) = |\psi_2\rangle\langle \psi_2|$, where $|\psi_1\rangle = (|S_1\rangle + |S_3\rangle)/\sqrt{2}$ and $|\psi_2\rangle = (|S_1\rangle - |S_3\rangle)/\sqrt{2}$. Clearly, these two states possess identical diagonal populations but opposite off-diagonal coherences.

Figure~\ref{h1} illustrates the dynamical evolution of the trace distance $D(t)$ between these two states under different dephasing strengths $\gamma_d$. In the absence of dephasing ($\gamma_d = 0$, green curves), $D(t)$ exhibits distinct behaviors depending on the specific non-Hermitian symmetry. However, once dephasing is introduced ($\gamma_d = 0.5$, orange curves), the trace distance decays rapidly to zero across all parameter regimes. More importantly, in the strong-dephasing limit ($\gamma_d = 200.0$, blue curves), $D(t)$ collapses to zero almost instantaneously at the onset of the evolution. This rapid collapse occurs because pure dephasing acts directly on the site populations and exponentially suppresses the off-diagonal coherence terms of the density matrix in the site basis. Since the distinguishability of $\hat{\rho}_1$ and $\hat{\rho}_2$ is entirely encoded in these coherences, the dephasing noise inevitably and rapidly washes away any initial difference between them.

\subsection{Initial-state dependent dephasing effects on population information: Analytical treatment and numerical results}

Having established that pure dephasing rapidly erases off-diagonal coherences, we now ask how it affects states that possess different initial spatial populations. In this subsection, we combine simplified analytical treatments with complete numerical simulations based on Eq.~\eqref{eq:master} to reveal that the spatial distribution of the initial state plays a decisive role in shaping the dephasing effects. Interestingly, we find that for the same set of initial states, systems across different non-Hermitian phases exhibit remarkably similar dynamical trends upon increasing the dephasing strength.
%In this subsection, we combine simplified analytical treatments with complete numerical simulations based on Eq.~\eqref{eq:master} to reveal the intricate effects of dephasing. In particular, we show both analytically and numerically that the initial state plays a major role in shaping these dephasing effects. For the same set of initial states, systems in different phases exhibit similar dynamical trends upon increasing the dephasing strength. 

\subsubsection{Equation of motion in the strong dephasing limit}
While the quantum master equation introduced in Eq.~\eqref{eq:master} provides a formally complete description of the system's dynamics in the presence of pure dephasing, its highly nonlinear structure and the intricate coupling between populations (diagonal elements) and coherences (off-diagonal elements) present significant challenges for obtaining direct analytical insights. To resolve the underlying physical mechanisms and to systematically understand how dephasing competes with non-Hermiticity, it is instructive to first examine the limiting regime of strong pure dephasing with $\gamma_d \to \infty$. In this limit, the rapid decay of quantum coherences allows us to decouple the fast and slow degrees of freedom, simplifying the full master equation into a set of effective, classically behaved rate equations for the diagonal elements. This simplification provides a tractable analytical framework to investigate initial-state-dependent dephasing effects that manifest in the strong-dephasing limit and enables us to elucidate the origin of this initial-state dependence. 

Specifically, in the strong dephasing limit, the dephasing rate $\gamma_d$ constitutes the dominant energy scale, significantly exceeding both the hopping parameters $|\eta_1|, |\eta_2|$ and the non-Hermitian strength $|\gamma|$. In this regime, the off-diagonal elements $\rho_{nm}(t)=\langle n|\hat{\rho}(t)|m\rangle\;(n\neq m)$ decay rapidly to zero on a extremely short timescale; For notational simplicity, we will use the shorthand $|n\rangle \equiv |S_n\rangle$ throughout this subsection. Consequently, the system's density matrix can be effectively approximated by a purely diagonal form
\begin{equation}
\hat{\rho}(t) \simeq \operatorname{diag}(\rho_{11}(t), \rho_{22}(t), \rho_{33}(t), \rho_{44}(t)). \label{eq:diagonal_approx}
\end{equation}
Here, $\rho_{nn}(t)=\langle n|\hat{\rho}(t)| n\rangle$ ($n=1,2,3,4$). $\operatorname{diag}(\cdot)$ denotes a diagonal matrix. Under this approximation, the contributions of the off-diagonal elements to the master equation emerge as higher-order infinitesimals  and can be safely neglected. Thus, understanding the dynamical evolution is reduced to solving the equations of motion solely for the diagonal populations. For simplicity, the time dependence is suppressed hereafter.

Strictly speaking, the exact time evolution of the diagonal element $\rho_{nn}$ is obtained by projecting Eq.~\eqref{eq:master} onto the local basis
\begin{equation}
\begin{split}
    \frac{d}{dt}\rho_{nn} &= \underbrace{-i\langle n|[\hat{H}\hat{\rho} - \hat{\rho}\hat{H}^{\dagger}]|n\rangle}_{A_n} \\
    &\quad + \underbrace{\langle n|\sum_k\mathcal{D}_k[\hat{\rho}]|n\rangle}_{B_n} \\
    &\quad \underbrace{- i\operatorname{Tr}\left[\hat{\rho}(\hat{H}^{\dagger} - \hat{H})\right]\rho_{nn}}_{C_n}.
\end{split}
\label{eq:diag_evolution}
\end{equation}
To obtain the correct equations of motion for the diagonal populations in the strong-dephasing limit, we examine the three terms $A_n$, $B_n$ and $C_n$ in turn.

We first focus on the term $A_n$. We expand it by inserting a complete set of states $\sum_m |m\rangle\langle m| = \mathrm{I}$,
\begin{equation}
    A_n = -i \sum_m \left( H_{nm}\rho_{mn} - \rho_{nm}H_{mn}^* \right).
\end{equation}
Since the Hamiltonian $\hat{H}$ contains Hermitian hopping and non-Hermitian on-site terms, its matrix elements satisfy $H_{nm}=H_{mn}$ for $n \neq m$, and $H_{nn} = i\gamma_n$ (We have $\gamma_{1,3}=-\gamma_{2,4}=\gamma$). Separating the $m=n$ term from the summation yields
\bea
   A_n &=& -i \left( i\gamma_n \rho_{nn} - \rho_{nn}(-i\gamma_n) \right) - i \sum_{m \neq n} H_{nm}(\rho_{mn} - \rho_{nm})\nonumber \\
    &=& 2\gamma_n \rho_{nn} - i \sum_{m \neq n} H_{nm}(\rho_{mn} - \rho_{nm}).\label{eq:An_expanded}
\eea
To evaluate the remaining summation, we must examine the off-diagonal elements $\rho_{nm}$ ($n \neq m$). The evolution of the off-diagonal elements is governed by
\begin{equation}\label{eq:144}
\begin{split}
    \frac{d}{dt}\rho_{nm} =& -i \sum_m (H_{nm}\rho_{mm} - \rho_{nm}H_{mm}^*) \\
    & - i\operatorname{Tr}\bigl[\hat{\rho}(\hat{H}^{\dagger} - \hat{H})\bigr]\rho_{nm} \\
    & - \gamma_d \rho_{nm}.
\end{split}
\end{equation}
In the strong dephasing limit of $\gamma_d \to \infty$, the off-diagonal elements almost instantaneously reach their stationary values with $\frac{d}{dt}\rho_{nm} \simeq 0$. Neglecting higher-order terms and retaining only the dominant populations in Eq. (\ref{eq:144}), we obtain
\begin{equation}
    0 \simeq -i H_{nm}(\rho_{mm} - \rho_{nn}) - \gamma_d \rho_{nm}.
\end{equation}
We thus obtain $\rho_{nm} \simeq -i \frac{H_{nm}}{\gamma_d} (\rho_{mm} - \rho_{nn})$. Inserting this expression back into the sum in Eq.~\eqref{eq:An_expanded}, we have
\bea
    -i H_{nm}(\rho_{mn} - \rho_{nm})
    &=& -i H_{nm} \biggl[ -i \frac{H_{mn}}{\gamma_d}(\rho_{nn} - \rho_{mm}) \nonumber\\
    && - \biggl( -i \frac{H_{nm}}{\gamma_d}(\rho_{mm} - \rho_{nn}) \biggr) \biggr] \nonumber\\
    &=& -i H_{nm} \biggl[ -2i \frac{H_{nm}}{\gamma_d}(\rho_{nn} - \rho_{mm}) \biggr] \nonumber\\
    &=& \frac{2 H_{nm}^2}{\gamma_d} (\rho_{mm} - \rho_{nn}).
\eea
Here, we note that off-diagonal elements $\{H_{nm}\} (n\neq m)$ in our system are real. Thus, the term $A_n$ is evaluated as
\begin{equation}
    A_n = 2\gamma_n \rho_{nn} + \sum_{m \neq n} \frac{2 H_{nm}^2}{\gamma_d} (\rho_{mm} - \rho_{nn}). \label{eq:An_2}
\end{equation}

We then turn to the terms $B_n$ and $C_n$. For the local pure dephasing with $\hat{L}_n = |n\rangle\langle n|$, the projected dissipator reads
\bea
    \langle n|\mathcal{D}_n[\hat{\rho}]|n\rangle &=& \gamma_d \Big(\langle n|n\rangle\langle n|\hat{\rho}|n\rangle\langle n|n\rangle \nonumber\\
    &&-\frac{1}{2}\langle n|\{|n\rangle\langle n|, \hat{\rho}\}|n\rangle \Big)=0.
\label{eq:Bn_1}
\eea
We thus infer that $B_n = 0$. This confirms that pure dephasing does not induce population transfer between sites. Lastly, to evaluate $C_n$, we first note that $S \equiv \operatorname{Tr}\left[\hat{\rho}(\hat{H}^{\dagger} - \hat{H})\right]=\sum_m \rho_{mm} (H_{mm}^* - H_{mm}) = -2i\sum_m \rho_{mm} \operatorname{Im}(H_{mm})$. Since $\operatorname{Im}(H_{mm}) = \gamma_m$, we can rewrite $S$ as
\begin{equation}
S = -2i\gamma(\rho_{11} - \rho_{22} + \rho_{33} - \rho_{44}) \equiv -2i\gamma M, \label{eq:S_2}
\end{equation}
where we have introduced the population imbalance parameter $M = \rho_{11} - \rho_{22} + \rho_{33} - \rho_{44}$. Consequently, the correction term is formulated as
\begin{equation}
C_n = -i S \rho_{nn} = -2\gamma M \rho_{nn}. \label{eq:Cn}
\end{equation}

By substituting Eqs.~\eqref{eq:An_2} and \eqref{eq:Cn} into Eq.~\eqref{eq:diag_evolution} and noting $B_n=0$, we obtain the simplified equation of motion for the population element $\rho_{nn}$ in the strong dephasing limit,
\begin{equation}
\frac{d}{dt}\rho_{nn} = 2\gamma_n \rho_{nn} + \sum_{m \neq n} \frac{2 H_{nm}^2}{\gamma_d} (\rho_{mm} - \rho_{nn}) - 2\gamma M \rho_{nn}. \label{eq:complete_diag}
\end{equation}
This equation of motion keeps the leading order term in $1/\gamma_d$. Defining the effective hopping rates $\kappa_1 = 2\eta_1^2/\gamma_d$ and $\kappa_2 = 2\eta_2^2/\gamma_d$, we obtain the complete set of coupled equations of motion for the four diagonal elements 
\begin{align}
    \dot{\rho}_{11} &= 2\gamma\rho_{11}(1 - M) + \kappa_1(\rho_{22} - \rho_{11}), \label{eq:rho_aa} \\[2pt]
    \begin{split}
    \dot{\rho}_{22} &= -2\gamma\rho_{22}(1 + M) + \kappa_1(\rho_{11} - \rho_{22}) \\
    &\quad + \kappa_2(\rho_{33} - \rho_{22}),
    \end{split} \label{eq:rho_bb} \\[2pt]
    \begin{split}
    \dot{\rho}_{33} &= 2\gamma\rho_{33}(1 - M) + \kappa_2(\rho_{22} - \rho_{33}) \\
    &\quad + \kappa_1(\rho_{44} - \rho_{33}),
    \end{split} \label{eq:rho_cc} \\[2pt]
    \dot{\rho}_{44} &= -2\gamma\rho_{44}(1 + M) + \kappa_1(\rho_{33} - \rho_{44}). \label{eq:rho_dd}
\end{align}
For clarity, we recall that Eqs.~\eqref{eq:rho_aa}-\eqref{eq:rho_dd} are obtained in the strong-dephasing limit, where off-diagonal coherences decay rapidly, reducing the evolution dynamics of $\hat{\rho}(t)$ to equations of motion for the diagonal populations that remain applicable at any finite times. 
%This does not imply, however, that Eqs.~\eqref{eq:rho_aa}--\eqref{eq:rho_dd} are applicable only to incoherent evolutions. In fact, we emphasize that these equations remain equally valid for systems initialized in both coherent and incoherent states. Indeed, as we will show later, the trace distance whose dynamics comes solely from the state would exhibit similar dynamical behaviors in systems initialized in both coherent and incoherent states in the strong-dephasing limit, all of which are faithfully captured by these equations. We remark that these equations can be applied to analyze the dynamical behaviors of other state-based information-theoretic quantities such as the Frobenius distance~\cite{Ma.25.PRB}.

In the strong-dephasing regime, the effective dynamics described by Eqs.~\eqref{eq:rho_aa}-\eqref{eq:rho_dd} acquire a clear physical interpretation in terms of population flow between sites with alternating gain and loss. The terms on the right-hand-side of Eqs.~\eqref{eq:rho_aa}--\eqref{eq:rho_dd} that are proportional to the non-Hermiticity strength $\gamma$ generate a pronounced local distinction: the local populations at gain sites are amplified, while those at loss sites experience suppression. Consequently, the long-time relaxation becomes highly sensitive to the initial state configuration. The remaining terms on the right-hand-side of Eqs.~\eqref{eq:rho_aa}-\eqref{eq:rho_dd}, which are induced by dephasing, tend to suppress spatial transport of population between sites, since the effective hopping rates $\kappa_1 = 2\eta_1^2/\gamma_d$ and $\kappa_2 = 2\eta_2^2/\gamma_d$ scale inversely with the dephasing strength.

This set of equations [Eqs.~\eqref{eq:rho_aa}--\eqref{eq:rho_dd}] serves as the theoretical basis for our subsequent identification of initial-state-dependent dephasing effects. That being said, the highly nonlinear nature of these equations--arising from the population imbalance parameter $M$ that originates from the correction term $C_n$--implies a strong sensitivity to initial conditions and presents a challenge for direct analytical treatment. To circumvent this nonlinearity, we draw inspiration from the formal solution of the master equation in Eq.~(\ref{xingshi}), where the correction term manifests as the normalization factor in the denominator. Analogously, we introduce a set of unnormalized populations $x_n$ such that the physical populations are subsequently recovered via instantaneous normalization
\begin{equation}
    \rho_{nn} = \frac{x_n}{\sum_{k=1}^4 x_k}.
    \label{eq:normalization_x}
\end{equation}
By factoring out the correction terms (i.e., the $-2\gamma M\rho_{nn}$ terms in Eqs.~\eqref{eq:rho_aa}--\eqref{eq:rho_dd}), the unnormalized populations $\{x_n\}$ obey a purely linear set of rate equations
\begin{align}
    \dot{x}_1 &= 2\gamma x_1 + \kappa_1(x_2 - x_1), \label{eq:x1_full} \\
    \dot{x}_2 &= -2\gamma x_2 + \kappa_1(x_1 - x_2) + \kappa_2(x_3 - x_2), \label{eq:x2_full} \\
    \dot{x}_3 &= 2\gamma x_3 + \kappa_2(x_2 - x_3) + \kappa_1(x_4 - x_3), \label{eq:x3_full} \\
    \dot{x}_4 &= -2\gamma x_4 + \kappa_1(x_3 - x_4). \label{eq:x4_full}
\end{align}
We have numerically verified that Eqs.~(\ref{eq:x1_full})-(\ref{eq:x4_full}) yield exactly the same population results at any time as Eqs.~\eqref{eq:rho_aa}--\eqref{eq:rho_dd} upon normalization according to Eq.~(\ref{eq:normalization_x}).

The linearized equations of motion in Eqs.~(\ref{eq:x1_full})-(\ref{eq:x4_full}) allow us to clearly isolate the competition between non-Hermiticity-induced local amplification/decay and dephasing-induced spatial diffusion in the strong-dephasing limit, with the latter can be regarded as perturbations since $\gamma \gg \kappa_{1,2}$. 

Specifically, since the effective hopping rates $\kappa_1 = 2\eta_1^2/\gamma_d$ and $\kappa_2 = 2\eta_2^2/\gamma_d$ scale inversely with the dephasing strength, the environmental dephasing strongly inhibits spatial transport. Nevertheless, residual population transfer between neighboring sites persists through these effective hopping channels. Simultaneously, the non-Hermitian gain and loss terms generate a pronounced directional amplification mechanism: the gain sites ($a$ and $c$) exponentially amplify incoming populations, whereas the loss sites ($b$ and $d$) exponentially suppress them. Consequently, the long-time dynamics are no longer governed by symmetric diffusion, but rather by a state-selective population flow toward the gain subspace. This mechanism can be directly inferred from Eqs.~(\ref{eq:x1_full})--(\ref{eq:x4_full}).

For an initial excitation located at the loss site $b$, the population is simultaneously transferred to its two neighboring gain sites $a$ and $c$ through the effective hopping channels $\kappa_1$ and $\kappa_2$. Once a small amount of probability reaches either gain site, the non-Hermitian amplification immediately enhances it exponentially. As a result, the probability initially localized at $b$ rapidly flows into the gain subspace spanned by $a$ and $c$. Similarly, an excitation initialized at the loss site $d$ predominantly flows toward the adjacent gain site $c$. Since $d$ is connected only to $c$, the probability leakage becomes effectively unidirectional in the strong-dephasing regime.

In contrast, populations initially prepared at the gain sites $a$ or $c$ exhibit qualitatively different behavior. Although weak leakage toward neighboring loss sites persists through $\kappa_{1,2}$, the exponential amplification at the original gain site overwhelmingly dominates the dynamics. After normalization, the physical state therefore remains macroscopically pinned to its initial gain location. This asymmetric population flow constitutes the microscopic origin of the initial-state-dependent dephasing effects observed in our numerical simulations.

In the following, we elucidate the initial-state-dependent dephasing effects by examining the dynamics of the trace distance $D(t)$ under different initial-state configurations with varying dephasing strengths. Before presenting the numerical results, we clarify the rationale for our choice of initial states for illustration. We have performed extensive numerical simulations across a wide range of initial conditions and found that, in the present setup, all observed dephasing effects fall into three distinct categories based on the behaviors of the trace distance in the strong-dephasing limit: accelerated decay, transient partial and complete stabilization. To clearly illustrate these generic dynamical trends, we therefore select representative initial states that capture the essence of each category. Moreover, for each chosen initial state, we employ the linearized rate equations in Eqs.~(\ref{eq:x1_full})--(\ref{eq:x4_full}) to analytically understand the numerical observations in the strong-dephasing limit, thereby providing a consistent theoretical underpinning for our results. We stress that Eqs.~(\ref{eq:x1_full})--(\ref{eq:x4_full}) provide a universal baseline to constrain behaviors only in the strong-dephasing limit; the detailed dynamical behaviors of the trace distance for finite dephasing strengths vary between initial states.

\begin{figure}[t!]
    \centering
    \includegraphics[width=1.0\linewidth]{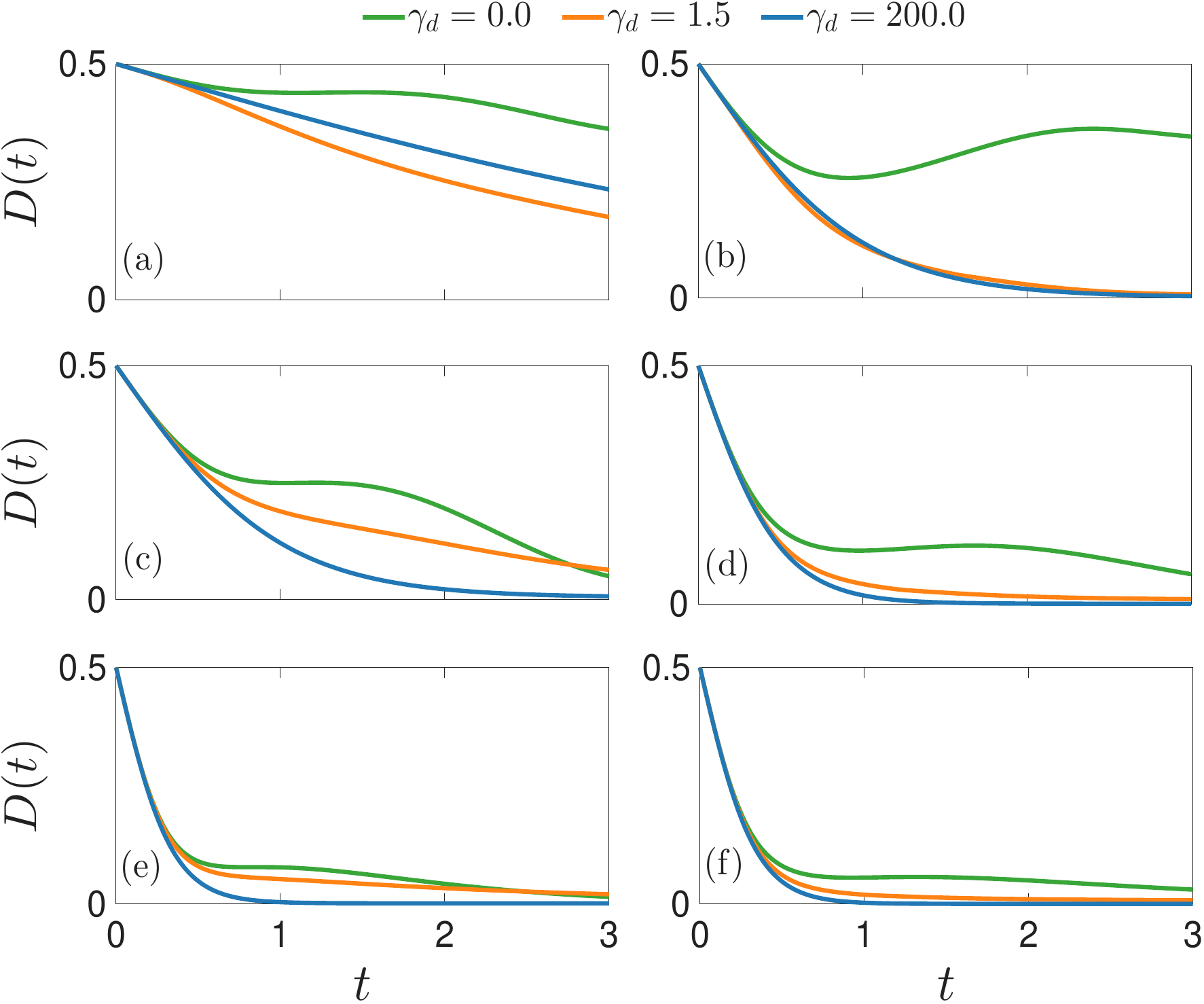}
    \caption{Dynamics of the trace distance $D(t)$ with varying dephasing strength $\gamma_d$ obtained by directly evolving Eq.~(\ref{eq:master}). (a) and (b) correspond to the $\mathcal{PT}$-unbroken phase for $\gamma = 0.1$ and $\gamma = 0.5$, respectively. (c) and (d) correspond to the symmetry-broken phase for $\gamma = 0.5$ and $\gamma = 1.0$, respectively. (e) and (f) correspond to the $\mathcal{APT}$-unbroken phase for $\gamma = 1.5$. Other parameters are $\eta_1 = 0.5$, $\eta_2 = 1$ for the left panel, and $\eta_1 = 1$, $\eta_2 = 0.5$ for the right panel. The initial states are $\hat{\rho}_1(0) = |S_1\rangle\langle S_1|$ and $\hat{\rho}_2(0) = \frac{1}{2}|S_1\rangle\langle S_1| + \frac{1}{2}|S_2\rangle\langle S_2|$.}
    \label{A22}
\end{figure}

\subsubsection{Case I: Accelerating decay}
We first demonstrate that dephasing can accelerate the decay of the trace distance, thus playing a detrimental role as in Hermitian systems. We showcase this phenomenon by harnessing the fact that local population at the gain site gets amplified while strong dephasing suppresses the spatial transport of population. Specifically, we consider two initial states: a pure state $\hat{\rho}_1(0) = |S_1\rangle\langle S_1|$, which is initialized at the first gain site, and a mixed state $\hat{\rho}_2(0) = \frac{1}{2}|S_1\rangle\langle S_1| + \frac{1}{2}|S_2\rangle\langle S_2|$, which evenly distributes initial population between the adjacent gain and loss sites. In the following, we use superscripts $(1)$ and $(2)$ to denote population dynamics from initial states $\hat{\rho}_1(0)$ and $\hat{\rho}_2(0)$, respectively.

For the initial pure state $\hat{\rho}_1(0)$, the unnormalized population is initialized at the first gain site with $x_1^{(1)}(0) = 1$. In the strong-dephasing limit, we have $\gamma \gg \kappa_{1,2}$, the unnormalized population $x_1^{(1)}(t)$ at finite times experiences an exponential amplification, $x_1^{(1)}(t) \simeq e^{2\gamma t}$. Upon normalization Eq.~\eqref{eq:normalization_x}, the physical state remains locked at its initial configuration, $\hat{\rho}_1(t) \simeq \operatorname{diag}(1,0,0,0)$.

As for the initial mixed state $\hat{\rho}_2(0)$, the unnormalized populations are initialized as $x_1^{(2)}(0) = 0.5$ and $x_2^{(2)}(0) = 0.5$, while $x_3^{(2)}(0) = x_4^{(2)}(0) = 0$. Governed overwhelmingly by the non-Hermitian terms ($\gamma \gg \kappa_{1,2}$), Eqs.~(\ref{eq:x1_full}) and (\ref{eq:x2_full}) indicate that the initial unnormalized populations at the first two sites approximately evolve as
\begin{equation}
    x_1^{(2)}(t) \simeq 0.5 e^{2\gamma t}, \qquad x_2^{(2)}(t) \simeq 0.5 e^{-2\gamma t}.
\end{equation}
To justify the omission of the effective hopping terms in evaluating the physical state, we explicitly calculate the probability leakage to the third site, which is initially unoccupied. Noting that the population $x_3^{(2)}(t)$ at finite times is continuously driven by leakage from the second site, we substitute $x_2^{(2)}(t)$ into Eq.~\eqref{eq:x3_full} and obtain a first-order inhomogeneous differential equation
\begin{equation}
    \dot{x}_3^{(2)} \simeq 2\gamma x_3^{(2)} + \kappa_2 \left(0.5 e^{-2\gamma t}\right).
\end{equation}
Solving this equation with the integrating factor $e^{-2\gamma t}$ and the initial condition $x_3^{(2)}(0) = 0$ yields
\begin{equation}
    x_3^{(2)}(t) \simeq \frac{\kappa_2}{8\gamma} \left( e^{2\gamma t} - e^{-2\gamma t} \right).
\end{equation}
In the strong-dephasing regime where $\gamma \gg \kappa_{1,2}$, the factor $\kappa_2/\gamma$ acts as a higher-order infinitesimal. This leads to a vanishing physical population: $\rho_{33}^{(2)}(t) \sim \mathcal{O}(\kappa_2/\gamma) \to 0$. An analogous sequential suppression applies to the fourth site, yielding $\rho_{44}^{(2)}(t) \sim \mathcal{O}(\kappa_1 \kappa_2 / \gamma^2) \to 0$. Therefore, the normalized physical state $\hat{\rho}_2(t)$ is determined almost entirely by the competition between the primary local gain and loss at the first two sites: $\hat{\rho}_2(t) \simeq \operatorname{diag}(\rho_{11}^{(2)}(t), \rho_{22}^{(2)}(t), 0, 0)$, where
\begin{align}
    \rho_{11}^{(2)}(t) &\simeq \frac{x_1^{(2)}(t)}{x_1^{(2)}(t) + x_2^{(2)}(t)} = \frac{e^{4\gamma t}}{1 + e^{4\gamma t}}, \label{eq:logistic1} \\[6pt]
    \rho_{22}^{(2)}(t) &\simeq \frac{x_2^{(2)}(t)}{x_1^{(2)}(t) + x_2^{(2)}(t)} = \frac{1}{1 + e^{4\gamma t}}. \label{eq:logistic2}
\end{align}
As can be seen, $\hat{\rho}_2(t)$ approaches $\hat{\rho}_1(t)$ exponentially in time.

This exponential approach of the states directly implies an exponential decay of the trace distance in the strong-dephasing limit:
\begin{equation}
\begin{aligned}
    D(t) &= \frac{1}{2} \left( \left|1 - \frac{e^{4\gamma t}}{1+e^{4\gamma t}}\right| + \left|0 - \frac{1}{1+e^{4\gamma t}}\right| \right) \\
    &= \frac{1}{1+e^{4\gamma t}}.
\end{aligned}
\end{equation}
Thus, the trace distance decays exponentially from $0.5$ toward $0$. To verify this theoretical prediction, we present a set of dynamical results for the trace distance in Fig.~\ref{A22} obtained by solving Eq.~(\ref{eq:master}) for the initial states $\hat{\rho}_1(0) = |S_1\rangle\langle S_1|$ and $\hat{\rho}_2(0) = \frac{1}{2}|S_1\rangle\langle S_1| + \frac{1}{2}|S_2\rangle\langle S_2|$. Upon increasing the dephasing strength, we indeed observe an exponentially accelerated decay of the trace distance in all three phases.

\begin{figure}[t!]
    \centering
    \includegraphics[width=1\linewidth]{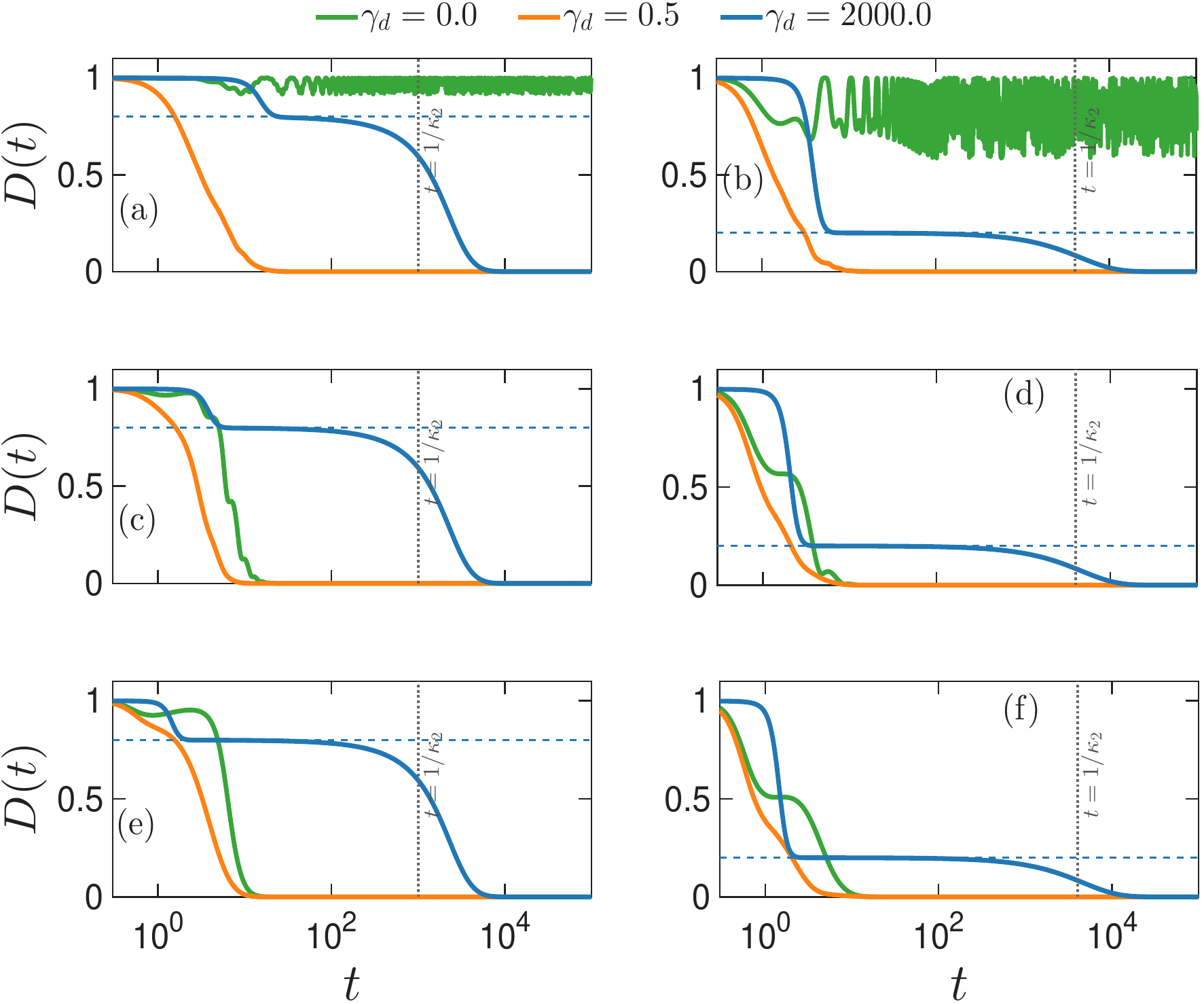}
    \caption{Dynamics of the trace distance $D(t)$ with varying dephasing strength $\gamma_d$ obtained by directly evolving Eq.~(\ref{eq:master}). (a) and (b) correspond to the $\mathcal{PT}$-unbroken phase for $\gamma = 0.1$ and $\gamma = 0.5$, respectively. (c) and (d) correspond to the symmetry-broken phase for $\gamma = 0.5$ and $\gamma = 1.0$, respectively. (e) and (f) correspond to the $\mathcal{APT}$-unbroken phase for $\gamma = 1.5$. Other parameters are $\eta_1 = 0.5$, $\eta_2 = 1$ for the left panel, and $\eta_1 = 1$, $\eta_2 = 0.5$ for the right panel. The initial states are $\hat{\rho}_1(0) = |S_1\rangle\langle S_1|$ and $\hat{\rho}_2(0) = |S_2\rangle\langle S_2|$. In each subfigure, the horizontal dashed line denotes the value of $D_{\rm pl}$ in Eq.~(\ref{eq:dpl1}) and the vertical dashed line marks the characteristic time $t=1/\kappa_2$.}
    \label{A222}
\end{figure}

\subsubsection{Case II: Transient partial stabilization}
We next show that increasing the dephasing strength does not always lead to accelerated decay of the trace distance toward zero, but can instead stabilize it to a finite value smaller than the initial value in the transient time regime. 

%\subsubsubsection{{\color{red}Incoherent initial states}}

%{\color{blue}Before analyzing specific initial states, we emphasize a general caveat that applies to all strong-dephasing results in this section and the next. In deriving the effective rate equations, we have neglected the terms proportional to $\kappa_1$ and $\kappa_2$ in the site equations, since they are small in the strong-dephasing limit. These terms, however, are not exactly zero and accumulate over a timescale of order $1/\kappa$.}

To illustrate this transient partial stabilization phenomenon, we consider a set of incoherent initial states in the $\{|S_i\rangle\}$ basis: $\hat{\rho}_1(0) = |S_1\rangle\langle S_1|$ and $\hat{\rho}_2(0) = |S_2\rangle\langle S_2|$, which are initialized at adjacent gain and loss sites, respectively.

For the initial state $\hat{\rho}_1(0)$, we previously showed that $\hat{\rho}_1(t) \simeq \operatorname{diag}(1,0,0,0)$. We therefore only need to analyze the dynamics starting from the initial state $\hat{\rho}_2(0) = |S_2\rangle\langle S_2|$, which implies that $x_2^{(2)}(0) = 1$ and all other unnormalized populations vanish. According to Eq.~\eqref{eq:x2_full}, its unnormalized population $x_2^{(2)}(t)$ in the strong-dephasing limit approximately evolves as
\begin{equation}
    x_2^{(2)}(t) \simeq e^{-2\gamma t}.
\end{equation}
%Crucially, the second site is coupled to two adjacent gain sites. Although the unnormalized populations $x_1^{(2)}$ and $x_3^{(2)}$ are initially zero, they become nonzero at later times due to leakage from $x_2^{(2)}$. Substituting $x_2^{(2)}(t) \simeq e^{-2\gamma t}$ into Eqs.~\eqref{eq:x1_full} and \eqref{eq:x3_full}, and neglecting the infinitesimal terms $-\kappa_1 x_1$ and $-(\kappa_1+\kappa_2) x_3$ ({\color{blue}since $2\gamma \gg \kappa_{1,2}$ in the strong-dephasing limit; these terms are negligible only at finite times, as they can accumulate in the exponent after normalization and eventually generate factors such as $e^{t/\gamma_d}$}), the linear equations for these adjacent gain sites simplify to first-order differential equations:
%\begin{align}
    %\dot{x}_1^{(2)} &\simeq 2\gamma x_1^{(2)} + \kappa_1 e^{-2\gamma t}, \\
    %\dot{x}_3^{(2)} &\simeq 2\gamma x_3^{(2)} + \kappa_2 e^{-2\gamma t}.
%\end{align}
Crucially, the second site is coupled to two adjacent gain sites. Although the unnormalized populations $x_1^{(2)}$ and $x_3^{(2)}$ are initially zero, they become nonzero at later times due to leakage from $x_2^{(2)}$. Substituting $x_2^{(2)}(t) \simeq e^{-2\gamma t}$ into Eqs.~\eqref{eq:x1_full} and \eqref{eq:x3_full}, and keeping the terms $-\kappa_1 x_1$ and $-(\kappa_1+\kappa_2) x_3$ that describe the back-action of the sites, the linear equations for these adjacent gain sites simplify to first-order differential equations:
\begin{align}
    \dot{x}_1^{(2)} &= (2\gamma - \kappa_1) x_1^{(2)} + \kappa_1 e^{-2\gamma t}, \\
    \dot{x}_3^{(2)} &= (2\gamma - \kappa_1 - \kappa_2) x_3^{(2)} + \kappa_2 e^{-2\gamma t}.
\end{align}
With the initial conditions $x_1^{(2)}(0) = x_3^{(2)}(0) = 0$, we obtain the following approximate solutions:
\begin{equation}
    x_1^{(2)}(t) = \frac{\kappa_1}{4\gamma - \kappa_1} \left(e^{(2\gamma - \kappa_1) t} - e^{-2\gamma t}\right),
\end{equation}
\begin{equation}
    x_3^{(2)}(t) = \frac{\kappa_2}{4\gamma - \kappa_1 - \kappa_2} \left(e^{(2\gamma - \kappa_1 - \kappa_2) t} - e^{-2\gamma t}\right).
\end{equation}
In the strong-dephasing limit $2\gamma \gg \kappa_{1,2}$, the prefactors simplify to $\kappa_1/(4\gamma)$ and $\kappa_2/(4\gamma)$, and the dominant long-time behavior is governed by the two slightly different growth rates $2\gamma - \kappa_1$ and $2\gamma - \kappa_1 - \kappa_2$. Their difference, $\kappa_2$, plays a crucial role after normalization.

Upon normalization $\rho_{nn} = x_n / \sum_k x_k$, the identical exponential growth terms $e^{2\gamma t}$ strictly cancel out. The system rapidly jumps into a stationary distribution which only has populations on the two gain sites
\begin{equation}
    \rho^{(2)}(t) \simeq \left( \frac{\kappa_1 e^{\kappa_2 t}}{\kappa_1 e^{\kappa_2 t} + \kappa_2}, 0, \frac{\kappa_2}{\kappa_1 e^{\kappa_2 t} + \kappa_2}, 0 \right).
    \label{ab}
\end{equation}
Calculating the trace distance between $\rho^{(1)}(t)$ and $\rho^{(2)}(t)$ gives
\begin{equation}\label{eq:443}
\begin{aligned}
    D(t) &= \frac{1}{2} \sum_n \left|\rho_{nn}^{(1)}(t) - \rho_{nn}^{(2)}(t)\right| \\
    &= \frac{\kappa_2}{\kappa_1 e^{\kappa_2 t} + \kappa_2}.
\end{aligned}
\end{equation}

We emphasize that Eq.~\eqref{eq:443} describes the dynamics after the initial fast transient, i.e., for $t \gg 1/\gamma_d$. At $t=0$, the two initial states are orthogonal, so $D(0)=1$. In the strong-dephasing regime, the population on the loss site $S_2$ decays on a timescale of order $1/\gamma_d$, and the trace distance rapidly drops from $1$ to the plateau value. Equation~\eqref{eq:443} captures the subsequent plateau and its slow decay, but not the initial rapid drop.

Equation~\eqref{eq:443} contains two distinct regimes. At times $t \ll 1/\kappa_2$ (strong dephasing), the exponential factor satisfies $e^{\kappa_2 t} \simeq 1$, and the trace distance approaches the plateau value
\begin{equation}\label{eq:dpl1}
    D_{\rm pl} = \frac{\kappa_2}{\kappa_1 + \kappa_2},
\end{equation}
which is fully determined by $\eta_{1,2}$ and independent of the phase. At much longer times $t \gg 1/\kappa_2$, the population on the first gain site dominates over that on the third gain site, and Eq.~\eqref{eq:443} reduces to
\begin{equation}
    D(t) \simeq \frac{\kappa_2}{\kappa_1} e^{-\kappa_2 t} \xrightarrow{t \gg 1/\kappa_2} 0.
\end{equation}
Therefore, the plateau at $D_{\rm pl}$ is a metastable quasi-steady value with lifetime of order $1/\kappa_2$, rather than the true $t \to \infty$ limit for finite dephasing strength. The true infinite-time limit is zero: for any finite $\gamma_d$, the two initial states eventually become indistinguishable because both approach the same gain-dominated steady distribution.

%{\color{blue}For the present initial states, The above derivation neglects the terms $-\kappa_1 x_1$ and $-(\kappa_1+\kappa_2) x_3$, which are small at short times but accumulate at longer times. These terms make the two gain modes grow with slightly different rates. As a result, after a timescale of order $1/\kappa_2$, the population on the first gain site dominates over that on the third gain site, and the distribution $\rho^{(2)}(t)$ eventually approaches $|S_1\rangle\langle S_1|$. Since $\rho^{(1)}(t)$ already approaches $|S_1\rangle\langle S_1|$, the trace distance ultimately decays to zero. Therefore, the derivation above only applies for a finite time.}

In Fig.~\ref{A222}, we present a set of numerical results for $D(t)$ under varying dephasing strength obtained by solving Eq.~(\ref{eq:master}) for the two initial states  $\hat{\rho}_1(0) = |S_1\rangle\langle S_1|$ and $\hat{\rho}_2(0) = |S_2\rangle\langle S_2|$. From the figure, we observe that increasing the dephasing strength first accelerates the decay of the trace distance compared with the dynamics in the closed system, but then stabilizes it to a finite plateau value. For times much longer than the plateau lifetime, the trace distance eventually decays to zero, in agreement with Eq.~\eqref{eq:443}. Notably, once $\eta_{1,2}$ are fixed, the plateau value of $D(t)$ remains constant across different phases, as expected. The horizontal dashed lines in Fig.~\ref{A222} show the analytic plateau value $D_{\rm pl}=\kappa_2/(\kappa_1+\kappa_2)$ obtained from the strong-dephasing analysis, while the vertical dashed lines indicate the characteristic time $t=1/\kappa_2$. The numerical curves in the strong-dephasing regime first undergo a rapid transient and then lie close to the horizontal dashed line for $t\ll 1/\kappa_2$, after which they slowly decay to zero. This directly confirms that the plateau is a metastable quasi-steady value with lifetime, rather than the true $t\to\infty$ limit.

\begin{figure}[t!]
    \centering
    \includegraphics[width=1\linewidth]{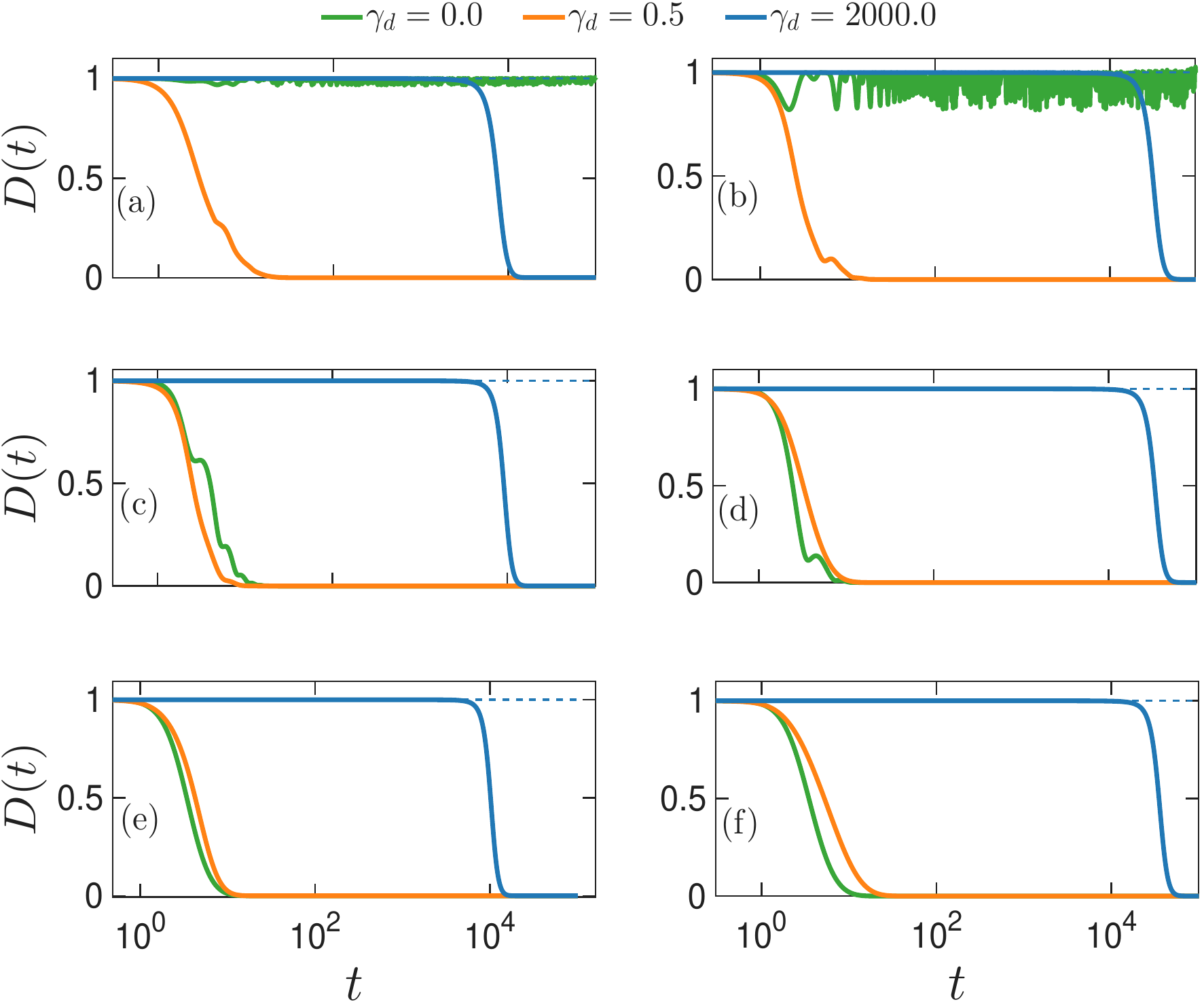}
    \caption{Dynamics of the trace distance $D(t)$ with varying dephasing strength $\gamma_d$ obtained by directly evolving Eq.~(\ref{eq:master}). (a) and (b) correspond to the $\mathcal{PT}$-unbroken phase for $\gamma = 0.1$ and $\gamma = 0.5$, respectively. (c) and (d) correspond to the symmetry-broken phase for $\gamma = 0.5$ and $\gamma = 1.0$, respectively. (e) and (f) correspond to the $\mathcal{APT}$-unbroken phase for $\gamma = 1.5$. Other parameters are $\eta_1 = 0.5$, $\eta_2 = 1$ for the left panel, and $\eta_1 = 1$, $\eta_2 = 0.5$ for the right panel. The initial states are $\hat{\rho}_1(0) = |S_1\rangle\langle S_1|$ and $\hat{\rho}_2(0)=|S_3\rangle\langle S_3|$. In
    each subfigure, the horizontal dashed line denotes the value of $D(t)=1$.}
    \label{A1}
\end{figure}

\subsubsection{Case III: Transient complete stabilization}
Interestingly, the previously discovered dephasing-induced transient stabilization can become complete in the sense that the evolution of the trace distance is entirely frozen, such that it remains at its initial value over a long time interval.

%\subsubsubsection{{\color{red}Incoherent initial states}}

We illustrate this transient complete stabilization of the trace distance using incoherent initial states $\hat{\rho}_1(0) = |S_1\rangle\langle S_1|$ and $\hat{\rho}_2(0) = |S_3\rangle\langle S_3|$, for which the system is initialized in two different gain sites.

As established, the initial state $\hat{\rho}_1(0)$ yields $\hat{\rho}_1(t) \simeq \operatorname{diag}(1,0,0,0)$ in the strong-dephasing limit. We now turn to the dynamics of $\hat{\rho}_2(t)$ initialized at the third site, which is a gain site. Retaining the $\kappa$ terms, the unnormalized population at this site evolves as
\begin{equation}
    x_3^{(2)}(t) \simeq e^{(2\gamma-\kappa_1-\kappa_2)t},
\end{equation}
To obtain the detailed form of $\hat{\rho}_2(t)$, we analyze the equations of motion for the nearby loss sites. Substituting 
$x_3^{(2)}(t) \simeq e^{(2\gamma-\kappa_1-\kappa_2)t}$
into Eqs.~\eqref{eq:x2_full} and \eqref{eq:x4_full} and 
keeping the dominant $\kappa$ terms,
we obtain
\begin{align}
    \dot{x}_2^{(2)} &\simeq -2\gamma x_2^{(2)} + \kappa_2 
     e^{(2\gamma-\kappa_1-\kappa_2)t}, \\
    \dot{x}_4^{(2)} &\simeq -2\gamma x_4^{(2)} + \kappa_1 
     e^{(2\gamma-\kappa_1-\kappa_2)t}.
\end{align}
Integrating these equations with zero initial populations yields
\begin{equation}
     x_2^{(2)}(t) \simeq \frac{\kappa_2}{4\gamma} e^{(2\gamma-\kappa_1-\kappa_2)t},
    \qquad
    x_4^{(2)}(t) \simeq \frac{\kappa_1}{4\gamma} e^{(2\gamma-\kappa_1-\kappa_2)t}.
\end{equation}
Because the total population is dominated by the source term at the third gain site ($\gamma \gg \kappa_{1,2}$), the normalized populations at the loss sites are suppressed: $\rho_{22} \sim \mathcal{O}(\kappa_2/\gamma)$ and $\rho_{44} \sim \mathcal{O}(\kappa_1/\gamma)$. In the transient plateau regime $t \ll 1/\kappa$ (under strong dephasing), the normalized density matrix therefore remains localized at the third site
\begin{equation}
    \hat{\rho}^{(2)}(t) \simeq \operatorname{diag}(0, 0, 1, 0).
\end{equation}
The corresponding trace distance in this transient plateau regime becomes
\begin{equation}
    D(t) \simeq \frac{1}{2}(|1-0| + |0-1|) = 1 = D(0).
\end{equation}
Thus, in the transient plateau regime, the dynamics of $D(t)$ is approximately frozen.

To confirm this theoretical expectation, in Fig.~\ref{A1} we present a set of numerical results for the trace distance $D(t)$ with varying dephasing strength $\gamma_d$, obtained by solving Eq.~(\ref{eq:master}) for the initial states $\hat{\rho}_1(0) = |S_1\rangle\langle S_1|$ and $\hat{\rho}_2(0) = |S_3\rangle\langle S_3|$. In the strong-dephasing regime, we clearly see that the dynamics of $D(t)$ is indeed strongly suppressed over a long plateau time, thereby confirming the transient complete stabilization phenomenon. For times much longer than the plateau lifetime, the trace distance eventually decays to zero. However, at small dephasing strengths, we identify a distinction between the $\mathcal{PT}$-unbroken and $\mathcal{APT}$-unbroken phases. In the former case, increasing the dephasing strength first transforms the persistent oscillations in the closed system into damped oscillations, thus playing a detrimental role. Only upon further increasing the dephasing strength does one observe a transient complete stabilization of the trace distance. In the latter case, by contrast, increasing the dephasing strength continuously suppresses the decay of the trace distance until 
transient complete stabilization is reached in the strong-dephasing limit.

\begin{figure}[b!]
    \centering
    \includegraphics[width=0.8\linewidth]{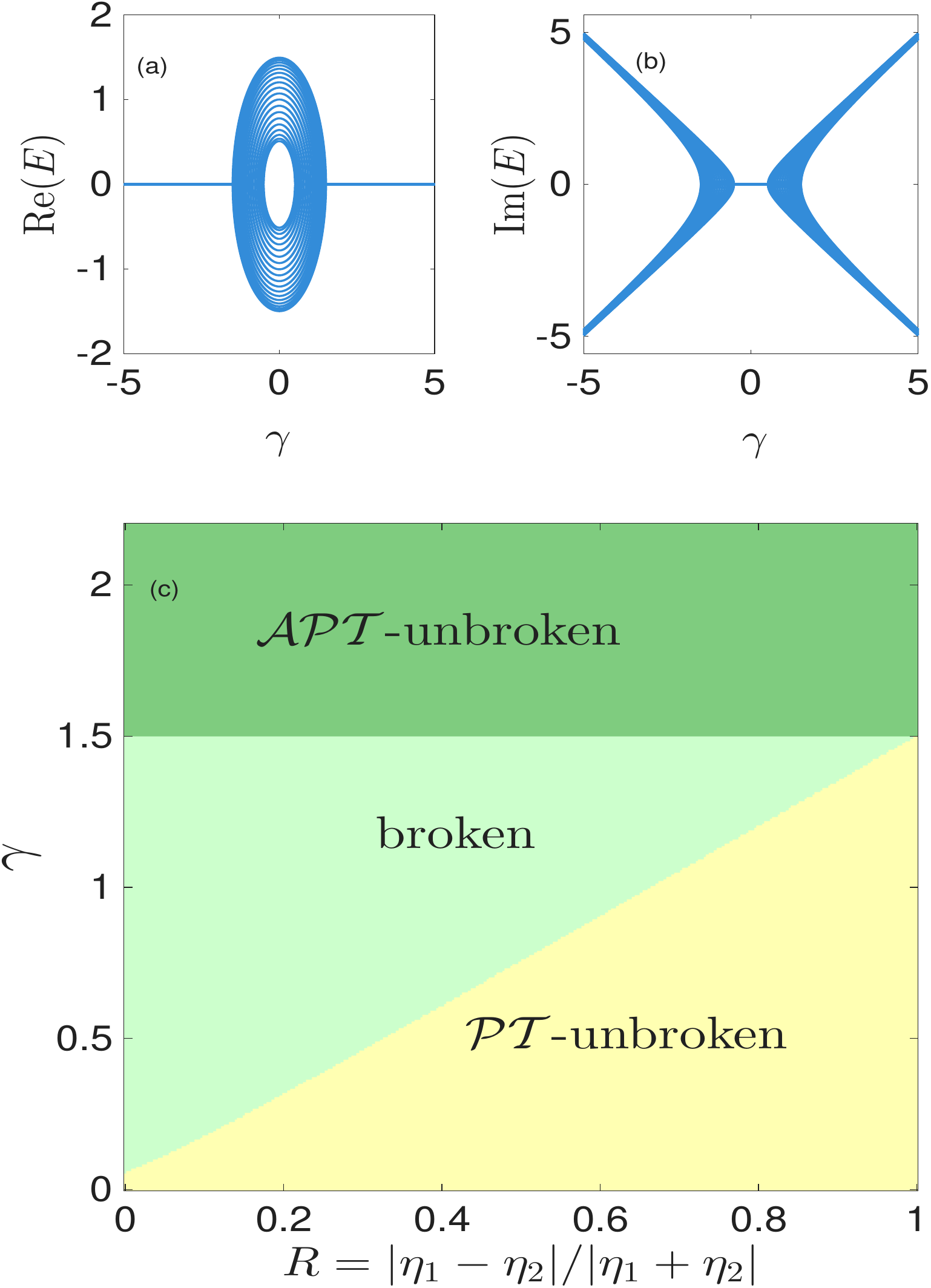}
    \caption{Spectral properties and phase diagram of the multi-site non-Hermitian SSH model with 10 unit cells. (a) Real and (b) imaginary parts of the energy eigenvalues as a function of $\gamma$ with $\eta_1=1$ and $\eta_2=0.5$. (c) Phase diagram in the $(\gamma, R)$ parameter space for $\eta_1 > \eta_2$, where $R = |\eta_1 - \eta_2| / (\eta_1 + \eta_2)$. The yellow, light green, and dark green shaded regions correspond to the $\mathcal{PT}$-unbroken, symmetry broken, and $\mathcal{APT}$-unbroken phases, respectively.}
    \label{M1}
\end{figure}

\begin{figure}[t!]
    \centering
    \includegraphics[width=1.0\linewidth]{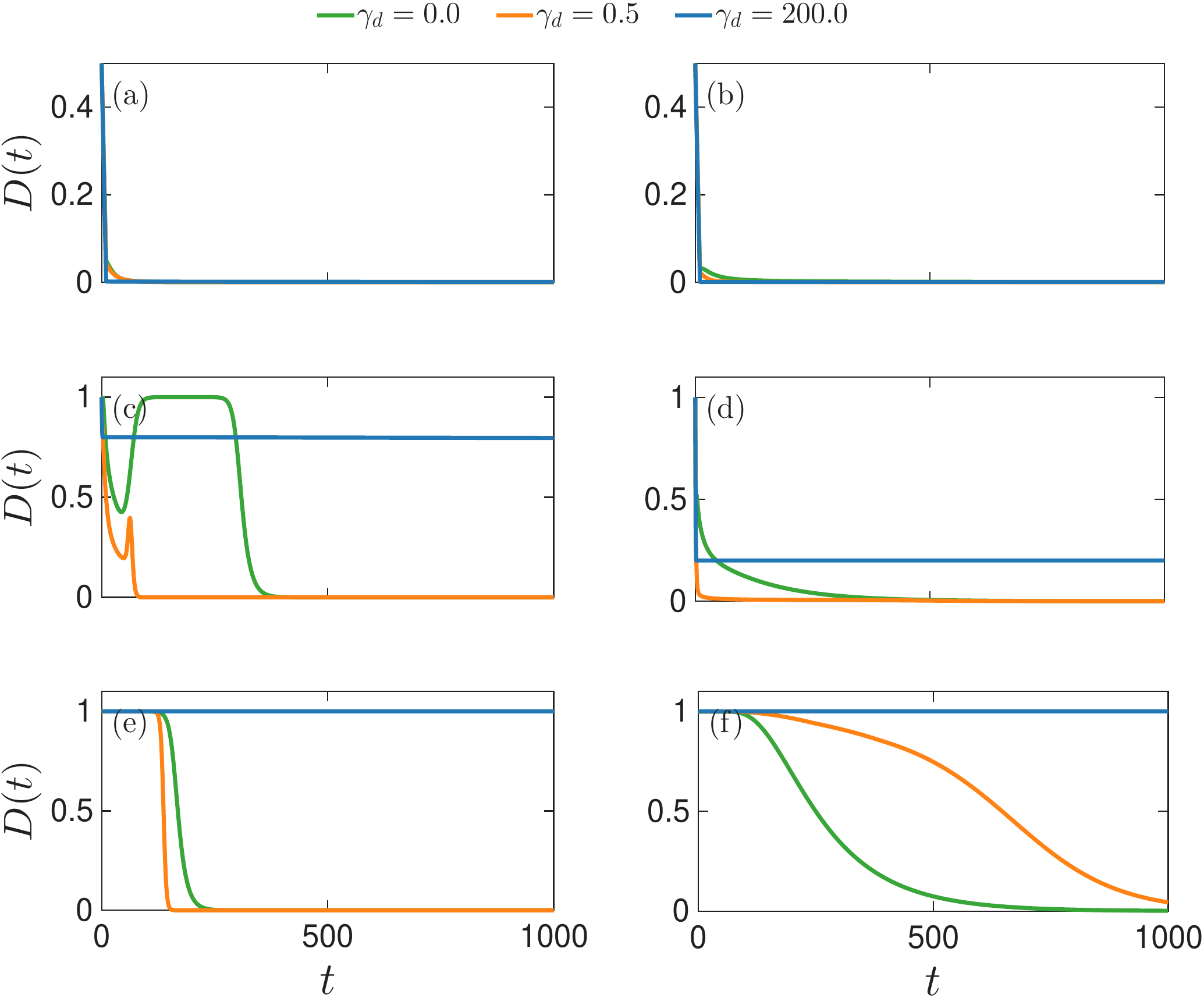}
    \caption{Time evolution of the trace distance $D(t)$ in a 40-site non-Hermitian SSH lattice system with varying pure dephasing strength $\gamma_d$. The three rows correspond to different initial state configurations: (a) and (b) $\hat{\rho}_1(0)=|S_1 \rangle\langle S_1|$ and $\hat{\rho}_2(0)=\frac12(|S_1\rangle\langle S_1|+|S_2\rangle\langle S_2|)$; (c) and (d) $\hat{\rho}_1(0)=|S_{17} \rangle\langle S_{17}|$ and $\hat{\rho}_2(0)=|S_{18}\rangle\langle S_{18}|$; (e) and (f) $\hat{\rho}_1(0)=|S_1 \rangle\langle S_1|$ and $\hat{\rho}_2(0)=|S_{39} \rangle\langle S_{39}|$. The system is in the $\mathcal{APT}$-unbroken phase with parameters $\gamma=2$. Other parameters are $\eta_1 = 0.5, \eta_2 = 1$ for the left panel, and $\eta_1 = 1, \eta_2 = 0.5$ for the right panel.}
    \label{A9}
\end{figure}

\section{Multi-site system}
\label{sec:multi_site}
To verify whether the observed phenomena persist in long chains, we extend our analysis to a multi-site non-Hermitian SSH model constructed by regarding the previously addressed four-site system as the unit cell. For demonstration purposes, we consider a one-dimensional chain under open boundary conditions consisting of $N = 10$ unit cells, amounting to a total of 40 sites. For each unit cell, we denote the corresponding annihilation operators as $\hat{a}_j, \hat{b}_j, \hat{c}_j$, and $\hat{d}_j$ for the $j$-th cell ($j = 1, \dots, N$). The total Hamiltonian is expressed as the sum of the Hermitian Su-Schrieffer-Heeger (SSH) part $\hat{H}_0$ and the non-Hermitian term $\hat{H}_\gamma$:
\begin{equation}
  \hat{H} = \hat{H}_0 + \hat{H}_\gamma,
\end{equation}
where
\begin{align}
  \hat{H}_0 &= \sum_{j=1}^{N} \bigl( \eta_1 \hat{a}_j^\dagger \hat{b}_j + \eta_2 \hat{b}_j^\dagger \hat{c}_j + \eta_1 \hat{c}_j^\dagger \hat{d}_j + \rm{H.c.} \bigr) \nonumber \\
  &\quad + \sum_{j=1}^{N-1} \bigl( \eta_2 \hat{d}_j^\dagger \hat{a}_{j+1} + \rm{H.c.} \bigr), \\
  \hat{H}_{\gamma} &= i\gamma \sum_{j=1}^{N} \bigl( \hat{a}_j^\dagger \hat{a}_j - \hat{b}_j^\dagger \hat{b}_j + \hat{c}_j^\dagger \hat{c}_j - \hat{d}_j^\dagger \hat{d}_j \bigr).
\end{align}
Here, $\eta_{1,2}$ remain the alternating inter-site coupling strengths, and $\gamma$ quantifies the strength of the balanced gain and loss processes uniformly applied across the entire lattice.

Parallel to the treatment of the four-site model, we restrict our analysis to the single-excitation subspace. For this 40-site system, the subspace is spanned by the Fock states $\{|S_1\rangle, |S_2\rangle, \dots, |S_{40}\rangle\}$, where $|S_n\rangle$ represents a single excitation localized on the $n$-th site of the entire chain, with all other sites empty (e.g., $|S_1\rangle = |100\dots0\rangle$, $|S_2\rangle = |010\dots0\rangle$, up to $|S_{40}\rangle = |000\dots1\rangle$). 

We numerically compute the energy spectrum of this Hamiltonian, as shown in Figs.~\ref{M1} (a) and (b). As expected, the spectral behavior resembles that of the four-site model. As the non-Hermitian strength $\gamma$ increases, the system undergoes a transition from a purely real spectrum to a purely imaginary one. The corresponding phase diagram is presented in Fig.~\ref{M1} (c), which clearly demarcates the distinct parameter regions of the $\mathcal{PT}$-unbroken, symmetry broken, and $\mathcal{APT}$-unbroken phases, confirming that the multi-site lattice preserves the fundamental symmetries of the minimal unit cell.

We then investigate the dynamical evolution of the trace distance $D(t)$ under different pure dephasing strengths $\gamma_d$, as shown in Fig.~\ref{A9} for incoherent initial states. We examine these dynamics by comparing two distinct parameter regimes: $\eta_1=0.5, \eta_2=1$ (left panel) and $\eta_1=1, \eta_2=0.5$ (right panel). Rather than observing a generic decay of quantum information due to the extensive size of the lattice, we find that the initial-state-dependent phenomena derived for the four-site model persist robustly in the 40-site chain. Specifically, Fig.~\ref{A9} (a) and (b), with $\hat{\rho}_1(0) = |S_1\rangle\langle S_1|$ and $\hat{\rho}_2(0) = \frac{1}{2}(|S_1\rangle\langle S_1| + |S_2\rangle\langle S_2|)$, reproduce the accelerated decay (Case I). In Fig.~\ref{A9} (c) and (d), we place the initial states deep within the bulk at adjacent gain and loss sites: $\hat{\rho}_1(0) = |S_{17}\rangle\langle S_{17}|$ and $\hat{\rho}_2(0) = |S_{18}\rangle\langle S_{18}|$. Even submerged in the 40-site lattice, strong dephasing isolates these sites from the rest of the chain, this leads to a partial stabilization of $D(t)$ (Case II), manifested as an intermediate-time plateau. 
%This plateau is the finite-chain counterpart of the partial stabilization found in the four-site model: strong dephasing strongly suppresses spatial transport, while the alternating gain and loss generate site-selective population flow, thereby forming a metastable quasi-steady plateau. 
Importantly, as in the four-site case, this plateau is not the true $t\to\infty$ steady state, but only a long-lived intermediate-time structure. For the 40-site chain, the time scale required to leave this plateau and eventually decay becomes extremely long, thus lying entirely outside the time window of our numerical simulations. This is consistent with the four-site results, which already explicitly revealed the plateau and its eventual decay.

Finally, Fig.~\ref{A9} (e) and (f) demonstrate complete stabilization (Case III). By initializing the states at gain sites, $\hat{\rho}_1(0) = |S_1\rangle\langle S_1|$ and $\hat{\rho}_2(0) = |S_{39}\rangle\langle S_{39}|$, strong dephasing completely severs any effective dynamic coupling across the extensive lattice, freezing the trace distance near its initial value. Similarly, the strong-dephasing curve here remains nearly frozen at its initial value over a long intermediate-time plateau. This frozen plateau should also be understood as metastable over the displayed time window; the eventual relaxation still occurs but lies outside the time range accessible in our simulations. Overall, these results demonstrate that the initial-state-dependent dephasing effects persist in chains of finite size beyond the minimal four-site system. More explicitly, the three dynamical categories--accelerated decay, transient partial and complete stabilization--persist qualitatively in the 40-site chain, showing that the underlying mechanism is robust against the system size. The main practical difference for the long chain is that the final decay after the intermediate-time plateau is pushed to much longer times, making it difficult to show the complete long-time relaxation within the numerical window. 

%{\color{red}To further verify that these features exist under coherent superposition states, we present in Fig.~\ref{xin40-2} the trace distance dynamics in the $\mathcal{APT}$-unbroken phase under coherent initial states. In the top row, we consider the initial states $\hat{\rho}_1(0) = |\psi_+\rangle\langle\psi_+|$ and $\hat{\rho}_2(0) = |\psi_-\rangle\langle\psi_-|$ with $|\psi_+\rangle = \frac{1}{\sqrt{2}}(|S_1\rangle + |S_2\rangle)$ and $|\psi_-\rangle = |S_1\rangle$, observing an accelerated decay of information similar to Case I. In the middle row, placing the superposition states into the bulk with $|\psi_+\rangle = \frac{1}{\sqrt{2}}(|S_{17}\rangle + |S_{19}\rangle)$ and $|\psi_-\rangle = \frac{1}{\sqrt{2}}(|S_{18}\rangle + |S_{20}\rangle)$ yields partial stabilization under strong dephasing (Case II). Lastly, the bottom row depicts the evolution for spatially separated superposition states, $|\psi_+\rangle = \frac{1}{\sqrt{2}}(|S_1\rangle + |S_2\rangle)$ and $|\psi_-\rangle = \frac{1}{\sqrt{2}}(|S_{39}\rangle + |S_{40}\rangle)$, which remarkably exhibits complete stabilization (Case III), fully suppressing the intrinsic decay.} Overall, these results demonstrate that the initial-state-dependent dephasing effects {\color{red}persist in chains of finite size beyond the minimal four-site system.}

\section{Conclusion}
\label{sec:conclusion}
In summary, we have investigated the dynamical properties of non-Hermitian SSH models with alternating gain and loss in real space under the influence of local pure dephasing. By employing a quantum Lindblad master equation to describe the open-system dynamics, we characterized the evolution of the trace distance across different symmetry phases, including the $\mathcal{PT}$-unbroken, $\mathcal{APT}$-unbroken, and symmetry-broken regimes.

Our key finding is that the dynamical response of the trace distance to dephasing is largely determined by the initial state rather than by the underlying Hamiltonian symmetries. By varying the initial states, we identified three distinct dynamical behaviors that become pronounced in the strong-dephasing limit:

(i) Accelerated decay, where dephasing acts as a conventional detrimental noise source;

(ii) Transient partial stabilization, where the trace distance asymptotically approaches a finite, nonzero value over a wide time window;

(iii) Transient complete stabilization, where the trace distance remains frozen at its initial value over a wide time window.

By deriving effective linearized rate equations for the diagonal populations in the strong-dephasing limit, we established that these behaviors originate from the competition between non-Hermiticity-induced local amplification/decay and dephasing-induced spatial transport of population, which is initial-state dependent. Finally, numerical simulations of a longer 40-site chain demonstrated that these initial-state-dependent dephasing effects survive multi-site systems. These findings clarify the interplay between dephasing and non-Hermiticity with spatial gain and loss, demonstrating that dephasing effect is typically initial-state dependent.

%In summary, we have systematically investigated the dynamical evolution of a non-Hermitian SSH model subject to local pure dephasing. To accommodate the unique normalization requirements inherent in non-Hermitian physics, we employed a nonlinear quantum master equation, which allowed us to unveil the microscopic mechanisms behind the anomalous relaxation slowdown in the strong dephasing limit.

%Our analysis demonstrates that the relaxation behavior is fundamentally determined by the symmetry phase of the system's Hamiltonian. In the four-site model, we identified that when the system is prepared in a pure state,a distinct protection mechanism in the APT-symmetric phase that induces a relaxation slowdown via a process physically independent of the traditional Quantum Zeno effect. In contrast, the relaxation slowdown observed in the PT-symmetric and PT-broken phases is found to be a direct consequence of the conventional QZE.

%Furthermore, we verified the scalability of this anomalous protection mechanism by extending our analysis from a four-site unit cell to a large-scale 40-site lattice. We confirm the robustness of this protection in multi-site systems. These findings provide profound insights into the interplay between non-Hermitian system, symmetry phases, and environmental.

\begin{acknowledgments}
This work is supported by the National Natural Science Foundation of China (Grant No. 12205179), the Shanghai Pujiang Program (Grant No. 22PJ1403900) and the Shanghai Science and Technology Innovation Action Plan (Grant No. 24LZ1400800).
\end{acknowledgments}

\appendix

\section{Symmetry of the non-Hermitian SSH model}\label{a:1}
In this Appendix, we provide the detailed algebraic derivation of the symmetries for the non-Hermitian SSH Hamiltonian defined in Eq.~\eqref{eq1}. We focus on the implementation of parity-time ($\mathcal{PT}$) and anti-parity-time ($\mathcal{APT}$) symmetries.

\subsection{Parity-Time Symmetry}
We first show that the Hamiltonian in Eq. (\ref{eq1}) exhibits $\mathcal{PT}$ symmetry. As defined in the main text, $\mathcal{T}$ represents the usual time-reversal operation, and $\mathcal{P}$ is the spatial inversion operator $\hat{\sigma}_x \otimes \hat{\sigma}_x$,
\begin{equation}
\mathcal{P} = \begin{pmatrix} 
0 & 0 & 0 & 1 \\ 
0 & 0 & 1 & 0 \\ 
0 & 1 & 0 & 0 \\ 
1 & 0 & 0 & 0 
\end{pmatrix}.
\end{equation}
One can easily check that the following relation holds,
\begin{equation}
\begin{aligned}
\mathcal{P} \hat{H}^* \mathcal{P} &= \begin{pmatrix} 0 & 0 & 0 & 1 \\ 0 & 0 & 1 & 0 \\ 0 & 1 & 0 & 0 \\ 1 & 0 & 0 & 0 \end{pmatrix}
\begin{pmatrix} -i\gamma & \eta_1 & 0 & 0 \\ \eta_1 & i\gamma & \eta_2 & 0 \\ 0 & \eta_2 & -i\gamma & \eta_1 \\ 0 & 0 & \eta_1 & i\gamma \end{pmatrix}
\begin{pmatrix} 0 & 0 & 0 & 1 \\ 0 & 0 & 1 & 0 \\ 0 & 1 & 0 & 0 \\ 1 & 0 & 0 & 0 \end{pmatrix} \\
&= \begin{pmatrix} i\gamma & \eta_1 & 0 & 0 \\ \eta_1 & -i\gamma & \eta_2 & 0 \\ 0 & \eta_2 & i\gamma & \eta_1 \\ 0 & 0 & \eta_1 & -i\gamma \end{pmatrix} = \hat{H}.
\end{aligned}
\end{equation}
namely, $\mathcal{P} \hat{H}^* \mathcal{P} = \hat{H}$, where $\hat{H}^*$ denotes the complex conjugate of $\hat{H}$. This relation is equivalent to the commutation relation $[\hat{H},\mathcal{PT}]=0$, implying that the system is $\mathcal{PT}$ symmetric. Hence, in the $\mathcal{PT}$-unbroken phase, the energy spectrum is purely real.

\subsection{Anti-Parity-Time Symmetry}
To see that the system also satisfies $\mathcal{APT}$ symmetry, we define the parity operator $\mathcal{P}'$ as
\begin{equation}
\mathcal{P}' = \text{diag}(1, -1, 1, -1),
\end{equation}
which yields the following transformation
\begin{equation}
\begin{aligned}
\mathcal{P}' \hat{H}^* \mathcal{P}' &= \begin{pmatrix} 1 & 0 & 0 & 0 \\ 0 & -1 & 0 & 0 \\ 0 & 0 & 1 & 0 \\ 0 & 0 & 0 & -1 \end{pmatrix}
\begin{pmatrix} -i\gamma & \eta_1 & 0 & 0 \\ \eta_1 & i\gamma & \eta_2 & 0 \\ 0 & \eta_2 & -i\gamma & \eta_1 \\ 0 & 0 & \eta_1 & i\gamma \end{pmatrix}
\begin{pmatrix} 1 & 0 & 0 & 0 \\ 0 & -1 & 0 & 0 \\ 0 & 0 & 1 & 0 \\ 0 & 0 & 0 & -1 \end{pmatrix} \\
&= \begin{pmatrix} -i\gamma & -\eta_1 & 0 & 0 \\ -\eta_1 & i\gamma & -\eta_2 & 0 \\ 0 & -\eta_2 & -i\gamma & -\eta_1 \\ 0 & 0 & -\eta_1 & i\gamma \end{pmatrix} = -\hat{H}.
\end{aligned}
\end{equation}
This relation is equivalent to the anti-commutation relation $\{\hat{H},\mathcal{P'T}\}=0$. Hence, the system indeed respects $\mathcal{APT}$ symmetry. In the $\mathcal{APT}$-unbroken phase, we expect a purely imaginary spectrum.

\section{Elucidating decaying patterns of the trace distance in the $\mathcal{APT}$-unbroken phase}
\label{app:multi_exponential}
In Sec.~\ref{TWO}~C of the main text, we unveiled rich decaying patterns of the trace distance in the $\mathcal{APT}$-unbroken phase of the closed non-Hermitian SSH model. Specifically, by simply tuning the initial-state pair $\{\hat{\rho}_1(0), \hat{\rho}_2(0)\}$, one can observe two kinds of dynamical trends of the trace distance $D(t)$: (i) Monotonic exponential decay, and (ii) non-monotonic decay. While the long-time asymptotic behavior is universally dictated by the slowest decay rate $\Gamma_1-\Gamma_2$, the rich dynamical features at shorter timescales demand a deeper look. This Appendix presents a thorough theoretical analysis for these transient dynamics.

We begin by recalling the time evolution of the normalized density matrix, Eq.~\eqref{eq:rho_t_exact} in the main text
\begin{equation}
\hat{\rho}(t) = \frac{\sum_{m,n}\rho_{mn}\,e^{-i(\lambda_m-\lambda_n)t}\,e^{(\Gamma_m+\Gamma_n)t}\,|\phi_m\rangle\langle\phi_n|}
{\mathrm{Tr}\bigl[\sum_{m,n}\rho_{mn}\,e^{-i(\lambda_m-\lambda_n)t}\,e^{(\Gamma_m+\Gamma_n)t}\,|\phi_m\rangle\langle\phi_n|\bigr]},
\label{eq:rho_exact}
\end{equation}
where $E_m = \lambda_m + i\Gamma_m$ ($m \in \{1,2,3,4\}$) are the complex eigenvalues of the non-Hermitian Hamiltonian $\hat{H}$. The coefficients $\rho_{mn} = \langle\chi_m|\hat{\rho}(0)|\chi_n\rangle$ represent the expansion components in the biorthonormal basis of $\hat{H}$ defined by $\langle\chi_m|\phi_n\rangle = \delta_{mn}$. 

In the $\mathcal{APT}$-unbroken phase, the energy spectrum is purely imaginary, meaning $\lambda_m = 0$ for all $m$, and thus $E_m = i\Gamma_m$. We order these imaginary parts in descending order: $\Gamma_1 > \Gamma_2 > \Gamma_3 > \Gamma_4$. Due to the fact that the system also supports $\mathcal{PT}$ symmetry, these eigenvalues are paired into $\pm$ doublets
\begin{equation}
\Gamma_4 = -\Gamma_1, \qquad \Gamma_3 = -\Gamma_2,
\label{eq:pt_pairing}
\end{equation}
which leaves only two independent positive parameters: $\Gamma_1 > \Gamma_2 > 0$.

To better isolate the relative decay rates, we factor out the dominant exponential term. By extracting $e^{2\Gamma_1 t}$ from both the numerator and denominator of Eq.~\eqref{eq:rho_exact}, we can define a set of non-negative decay rates
\begin{equation}
\Delta_m \equiv \Gamma_1 - \Gamma_m \ge 0.
\label{eq:Delta_def}
\end{equation}
Utilizing the pairing symmetry in Eq.~\eqref{eq:pt_pairing}, the explicit values of these rates are $\Delta_1 = 0$, $\Delta_2 = \Gamma_1 - \Gamma_2$, $\Delta_3 = \Gamma_1 + \Gamma_2$, and $\Delta_4 = 2\Gamma_1$. Consequently, the normalized density matrix simplifies to
\begin{equation}
\hat{\rho}(t) = \frac{\displaystyle\sum_{m,n=1}^{4}\rho_{mn}\,e^{-(\Delta_m+\Delta_n)t}\,|\phi_m\rangle\langle\phi_n|}
{\displaystyle\sum_{m,n=1}^{4}\rho_{mn}\,e^{-(\Delta_m+\Delta_n)t}\,N_{nm}},
\label{eq:rho_reduced}
\end{equation}
where $N_{nm} \equiv \langle\phi_n|\phi_m\rangle$ represents the overlap matrix of the right eigenstates. Crucially, because $\hat{H}$ is non-Hermitian, these eigenstates are generally non-orthogonal, meaning the off-diagonal elements are nonzero (i.e., $N_{nm} \neq \delta_{nm}$).

The full evolution in Eq.~\eqref{eq:rho_reduced} is governed by the sum of exponents $\Delta_m + \Delta_n$ across all 16 pairs of $(m,n) \in \{1,2,3,4\}^2$. Explicitly evaluating this matrix sum yields
\begin{equation}
\renewcommand{\arraystretch}{1.5}
\begin{array}{c|cccc}
\Delta_m+\Delta_n & n=1 & n=2 & n=3 & n=4 \\\hline
m=1 & 0 & \Gamma_1-\Gamma_2 & \Gamma_1+\Gamma_2 & 2\Gamma_1\\
m=2 & \Gamma_1-\Gamma_2 & 2(\Gamma_1-\Gamma_2) & 2\Gamma_1 & 3\Gamma_1-\Gamma_2\\
m=3 & \Gamma_1+\Gamma_2 & 2\Gamma_1 & 2(\Gamma_1+\Gamma_2) & 3\Gamma_1+\Gamma_2\\
m=4 & 2\Gamma_1 & 3\Gamma_1-\Gamma_2 & 3\Gamma_1+\Gamma_2 & 4\Gamma_1
\end{array}
\label{tab:Delta_sum}
\end{equation}
By grouping the identical entries in this matrix, we find that the 16 combinations collapse into exactly nine distinct non-negative decay rates, denoted as $\Omega_q$. Their corresponding multiplicities and specific $(m,n)$ pairings are summarized in Table~\ref{tab:nine_rates}.

\begin{table}[h]
\centering
\caption{The nine distinct decay rates $\Omega_q$ and their respective multiplicities. The sum of multiplicities ($1+2+1+2+4+2+1+2+1=16$) fully accounts for all $(m,n)$ pairs.}
\label{tab:nine_rates}
\renewcommand{\arraystretch}{1.25}
\begin{tabular}{c|c|l}
\hline
Rate $\Omega_q$ & Multiplicity & $(m,n)$ pairs \\\hline
$0$ & 1 & $(1,1)$\\
$\Gamma_1-\Gamma_2$ & 2 & $(1,2),(2,1)$\\
$2(\Gamma_1-\Gamma_2)$ & 1 & $(2,2)$\\
$\Gamma_1+\Gamma_2$ & 2 & $(1,3),(3,1)$\\
$2\Gamma_1$ & 4 & $(1,4),(4,1),(2,3),(3,2)$\\
$3\Gamma_1-\Gamma_2$ & 2 & $(2,4),(4,2)$\\
$2(\Gamma_1+\Gamma_2)$ & 1 & $(3,3)$\\
$3\Gamma_1+\Gamma_2$ & 2 & $(3,4),(4,3)$\\
$4\Gamma_1$ & 1 & $(4,4)$\\\hline
\end{tabular}
\end{table}

This mathematical decomposition offers a clear physical insight: a purely imaginary energy spectrum does not imply simple, single-exponential dynamics. Instead, due to the trace-normalization in the denominator, the state $\hat{\rho}(t)$ behaves as a non-linear, fractional superposition of nine independent real exponential decay terms:

(i) In the long-time regime of $t \to \infty$, all higher-rate channels rapidly die out. The system is universally dominated by the slowest available non-zero rate, $\Omega = \Gamma_1 - \Gamma_2$, consistent with the observations in the main text.

(ii) In the short- and intermediate-time regimes, the remaining eight transient channels ($\Omega_q > \Gamma_1 - \Gamma_2$) remain active. Because the choice of initial states drastically alters the initial expansion coefficients $\rho_{mn}$, it selectively activates or suppresses different decay channels. This multi-index competition between the nine channels is precisely what gives rise to the non-monotonic bumps, sharp dips, or purely monotonic decays observed in the trace distance.

%\bibliographystyle{ieeetr}
%\bibliography{Ref}


\end{document}